\documentclass[11pt]{article}
\usepackage[a4paper,top=2cm,bottom=2cm,left=2cm,right=2cm,marginparwidth=1.75cm]{geometry}

\usepackage{amsmath}
\usepackage{amssymb}
\usepackage{amsthm}

\usepackage{graphicx}
\usepackage{subcaption}
\usepackage{makecell}
\usepackage{xcolor}
\usepackage[colorlinks=true, allcolors=blue]{hyperref}
\usepackage[capitalise]{cleveref}

\usepackage{xurl}
\usepackage{booktabs}
\usepackage{graphicx}
\usepackage{authblk}
\usepackage{orcidlink}
\usepackage{lineno}
\usepackage{setspace}
\usepackage{enumitem}

\usepackage{comment}
\newtheorem{definition}{Definition}

\newtheorem{theorem}{Theorem}

\newcommand{\gendivergence}[3]{#1\left(#2 \mid \mid #3\right)}

\usepackage[bibstyle=numeric, citestyle=numeric-comp, sorting=none, isbn=false, url=false, eprint=false]{biblatex}
\title{From First Principles to Multi-scale Decomposition:\\Mutual Information as a Segregation Index}
\author[1,2,$\dagger$]{Rohit Sahasrabuddhe\,\orcidlink{0000-0002-2779-8310}}
\author[1,$\ddagger$]{Renaud Lambiotte\,\orcidlink{0000-0002-0583-4595}}
\affil[1]{Mathematical Institute, University of Oxford, Oxford, United Kingdom}
\affil[2]{Institute for New Economic Thinking, University of Oxford, Oxford, United Kingdom}
\affil[$\dagger$]{Corresponding author: rohit.sahasrabuddhe@maths.ox.ac.uk}
\affil[$\ddagger$]{Corresponding author: renaud.lambiotte@maths.ox.ac.uk}

\begin{document}
\maketitle
\begin{abstract}
    Segregation operates across demographic and spatial scales and requires careful and principled measurement. In this work, we derive mutual information from well-known general approaches to constructing segregation indices and interpret the mathematical properties that are unique to it. The properties relate mainly to its behaviour when demographic identities and spatial units are coarse-grained. Mutual information decomposes into the contributions of regions and supergroups, which reveals insights that would be inaccessible otherwise. We demonstrate this with an exploratory case study of residential ethnic segregation in England and Wales. Overall, we help researchers identify when mutual information is the only suitable segregation index and provide open source software to support multi-scale analysis.
\end{abstract}

\section{Introduction}

Measuring segregation across geography, demography, and time is critical for understanding persistent inequality and informing urban planning. Residential segregation is an important driver of unequal access to education, employment, housing, and public services. Neighbourhoods vary greatly in economic outcomes~\cite{Chetty2014, Sharkey2014} and have a strong effect on inter-generational mobility~\cite{Chetty2018a, Chetty2018b}. The increasing availability of high-resolution social network~\cite{Tth2021} and spatial mobility data~\cite{GreenbergRaanan2014, Wang2018, Huck2019, Moro2021} provides the opportunity to add more nuance to our understanding of the structure and dynamics of segregation.

How do we quantify segregation? Researchers have been developing segregation indices for decades~\cite{Duncan1955, White1986, Massey1988} and there is clearly no one-size-fits-all approach. In addition to interpretability, we can evaluate indices by whether they have desirable mathematical properties~\cite{James1985}. One such criterion is whether an index helps us understand the spatial and demographic scales of segregation. In this work, we use `scale' to refer to a level of aggregation: spatial units can be grouped into regions and demographic identities into supergroups. Segregation manifests across scales. While analyses at broad scales can reveal far-ranging inequalities, there is often important variation at finer scales. For instance, while African Americans, Hispanics, and Asians are all minorities with long histories of discrimination and segregation in the United States~\cite{Massey1990}, their experiences are far from identical~\cite{Denton1988}. There are more differences at even finer scales, such as country of origin~\cite{Iceland2008, Scopilliti2008, Iceland2014}. Identifying meaningful scales is akin to coarse-graining data while retaining essential structure~\cite{Chodrow2017}.

A segregation index that can be decomposed into the contributions of regions and supergroups allows us to disentangle how local processes aggregate into large-scale patterns~\cite{Reardon2004}. Such decomposability is a hallmark of information theory~\cite{Shannon1948, Cover2005}, which \textcite{Theil1967} introduced into segregation research. \textcite{Theil1971}'s information theory index is widely used to measure segregation across scales~\cite{Zoloth1976, Reardon2000, Fischer2004, Fiel2013, Lichter2015}. The information theory index is a normalisation of the index we study in this work: mutual information. Mutual information enjoys a stronger decomposability~\cite{Mora2011} and \textcite{Frankel2011} identify properties that are unique to rankings of regions by it.

How do we interpret mutual information as a segregation index? \textcite{Reardon2002} propose general approaches to constructing a multi-group segregation index as (a) the disproportionality of group proportions, (b) a diversity ratio, and (c) a measure of association\footnote{They recommend against averaging two-group indices since it ``\textit{may divorce an index from its substantive interpretation}"~\cite{Reardon2002}}. In \cref{sec:theory}, we derive mutual information from each of the approaches by identifying mathematical properties that are unique to it. Our contribution is to interpret these properties in the context of segregation, helping researchers decide when mutual information is the appropriate segregation index for their research question.

In \cref{sec:empirics}, we demonstrate the usefulness of mutual information's decomposition by investigating residential ethnic segregation in England and Wales. The increasing ethnic diversity in these countries has complex patterns of segregation across ethnicities and regions~\cite{Johnston2013, Johnston2015, Catney2016, Catney2023}, and segregation is a prominent topic of public discourse~\cite{BBCFarage, GuardianSuella, LBCJenrick}. Using high-resolution census data from 2021, we show that mutual information picks up on important structure across spatial and demographic scales.

\section{Mutual information from first principles} \label{sec:theory}

\subsection{Preliminaries}
We denote the set of $n$ demographic identities as $\mathcal{X}$ and the set of $m$ spatial units using $\mathcal{S}$. Each individual belongs to a single identity $x\in\mathcal{X}$ and spatial unit $s\in\mathcal{S}$. Our setting is `aspatial': we do not directly account for the proximity of spatial units~\cite{Reardon2004}. This does not preclude tasks such as regionalisation which use these indices from being spatially constrained~\cite{Chodrow2017}.

$X$ and $S$ are the random variables that take the value of an individual's demographic label and spatial unit respectively. From empirical data, we estimate the joint distribution $P_{X,S}$ such that $P_{X,S}\left( x,s\right)$ is the probability that an individual chosen uniformly at random has demographic identity $x$ and location $s$. The marginals $P_X$ and $P_S$ are the population distributions over demography and geography respectively. The demographic composition of $s$ is the conditional distribution $P_{X \mid S=s}$. We use $\Delta_n$ to denote the set of probability distributions over $n$ elements and $P \in \Delta_n$ to denote a generic distribution.

\subsection{Segregation as disproportionality}
In this approach, segregation is the mean demographic dissimilarity of spatial units from the population. How do we quantify the dissimilarity between two demographic compositions? Divergences are a wide class of options, requiring only that dissimilarity is 0 if they are identical and positive otherwise.

\begin{definition}
    A \textbf{divergence} is a function $f:\Delta_n \times \Delta_n \to [0, \infty]$ such that $\gendivergence{f}{P}{Q}=0$ if and only if $P=Q$.
\end{definition}

Using a divergence $f$, segregation $I$ is the mean divergence of a spatial unit's demographic composition from the population's. We weight the contribution of each unit by its population.

\begin{equation}
    I = \sum_{s \in \mathcal{S}} P_S(s) \; \gendivergence{f}{P_{X\mid S=s}}{P_X}
\label{eq:segregation_sum_over_s}
\end{equation}

Our starting point generalises \textcite[Eq. 3]{Reardon2002}'s approach while satisfying their disproportionality axiom: segregation is $0$ if all locations are demographically identical and positive otherwise. To narrow down the choice of $f$, let us consider potentially desirable properties for it to have.

\subsubsection{The mean-minimiser property}
Given the joint distribution $P_{X,S}$, what is the demographic composition of the population? Aggregating the observations of each spatial unit gives us the marginal $P_X$, which is what we used in \cref{eq:segregation_sum_over_s}. Abandoning this intuition for a moment, let us formulate the general task. Consider a set of $m$ distributions $\left\{ P_1,\,\dots,\,P_m \right\}$, with each $P_i\in\Delta_n$, and positive weights $w_1,\,\dots,\,w_m$ that sum to 1. How should we choose a single distribution to represent them? For a divergence $f$, a best representative $Q^*\in\Delta_n$ is a distribution that minimises the mean divergence from it.
\begin{equation}
     Q^* = \underset{Q \in \Delta_n}{\textnormal{argmin}} \; \sum_{i=1}^m w_i\, \gendivergence{f}{P_i}{Q}
\end{equation}

\begin{definition}
    A divergence $f$ has the \textbf{mean-minimiser property} if
    \begin{equation}
         \underset{Q \in \Delta_n}{\textnormal{argmin}} \; \sum_{i=1}^m w_i\, \gendivergence{f}{P_i}{Q} =  \sum_{i=1}^m w_i \,P_i
    \end{equation}
    uniquely.
\end{definition}

Note that the marginal is the mean of the conditionals.
\begin{equation}
    P_X = \sum_{s \in \mathcal{S}} P_{S}(s) \; P_{X \mid S=s}
\label{eq:marginal_expectation_conditionals}
\end{equation}
Thus, for the marginal to be the unique best representative, $f$ must have the mean-minimiser property. This aligns with how we aggregate observations and ensures that $I$ cannot be decreased by choosing a representative different from $P_X$. \textcite{Banerjee2005} proved that alongside some regularity conditions, this restricts $f$ to being a Bregman divergence~\cite{Bregman1967}.

\begin{definition}
    Let $\phi:\Delta_n \to \mathbb{R}$ be a strictly convex and differentiable function. The associated \textbf{Bregman divergence} is the function $f_\phi:\Delta_n \times \Delta_n^\circ \to [0, \infty )$ defined as
    \begin{equation}
        \gendivergence{f_\phi}{P}{Q} = \phi\left(P\right) - \phi\left(Q\right) - \langle \nabla \phi(Q) ,\; P-Q \rangle,
    \end{equation}
    where $\langle \cdot \mid  \cdot \rangle$ is the standard inner product in $\mathbb{R}^n$.
\end{definition}
Here, $\Delta_n^\circ$ is the set of distributions with full support. Many distance-like functions are Bregman divergences, such as squared Euclidean, squared Mahalanobis, and the Itakura-Saito distances.

\begin{theorem} (\textcite[Theorem 4]{Banerjee2005})
    For $n \geq 2$, consider a divergence $f$. If
    \begin{enumerate}[noitemsep, topsep=0pt]
        \item $f$ and its second derivatives with respect to $P$ ($\gendivergence{f_{P(i)P(j)}}{P}{Q}$ for $1 \leq i,j\leq n$) are all continuous and
        \item $f$ has the mean-minimiser property,
    \end{enumerate}
    then $f$ is a Bregman divergence $f_\phi$.
\end{theorem}

In the language of Bregman geometry, $I$ (\cref{eq:segregation_sum_over_s}) is Bregman information~\cite{Banerjee2004}. \textcite[Table S1]{Chodrow2017} points out that many segregation indices are Bregman informations. We note that while the mean-minimiser property is generally convenient, it may not be relevant when segregation is measured against a theoretical or normative distribution~\cite{Roberto2024}.

\subsubsection{The data processing property}\label{sec:dpi}

Segregation operates across scales, making it important to consider how $I$ behaves when spatial units or demographic identities are aggregated. Grouping induces a natural transformation of probability distributions: the probability of a group is the sum of the probabilities of its elements. How should coarse-graining demographic identities into supergroups affect divergence? Intuitively, ignoring the distinction between some identities should not increase the divergence. Generalising this to stochastic transformations brings us to the data processing property.

Consider a stochastic transformation represented by the $n \times n$ transition matrix $T$ where $T_{xy}$ is the probability that $x$ is mapped to $y$. We denote $TP$ as the distribution obtained by applying $T$ to $P$ such that $TP(y) = \sum_x P(x)\,T_{xy}$.
\begin{definition}
    A divergence $f$ has the \textbf{data processing property} if for any stochastic transformation,
    \begin{equation}
        \gendivergence{f}{TP}{TQ} \leq \gendivergence{f}{P}{Q}
    \end{equation}
    for all $P,Q \in \Delta_n$.
\end{definition}
Informally, the data processing property means that post-processing distributions does not increase divergence. \textcite{Jiao2014} show that for general $n$, the only Bregman divergence that satisfies the data processing property is Kullback-Leibler divergence~\cite{Kullback1951}.

\begin{definition}
    \textbf{Kullback-Leibler (KL) divergence} $d_\textnormal{KL}:\Delta_n\times\Delta_n\to[0,\infty]$ is given by
    \begin{equation}
        \gendivergence{d_\textnormal{KL}}{P}{Q} = \sum_{i=1}^n P(i)\, \log \frac{P(i)}{Q(i)}
    \end{equation}
    with the convention $0\log0=0$ for $P$ and $Q$ such that $P(i)=0$ if $Q(i)=0$. It is $\infty$ otherwise.
\end{definition}

\begin{theorem} (\textcite[Theorem 4]{Jiao2014})
    For $n\geq3$, if a Bregman divergence satisfies the data processing property, then it is KL divergence up to a non-negative multiplicative factor.
\label{thm:jiao}
\end{theorem}

We note that this version of the theorem is weaker than the original, which uses a weaker version of data processing. Since $I$ is a sum of divergences (\cref{eq:segregation_sum_over_s}), it scales with the same multiplicative factor applied to $d_\textnormal{KL}$. As all the transformations in our analysis are linear, we may set this constant to 1 without loss of generality. This leads to our first formulation of mutual information: the Bregman information of KL divergence.
\begin{equation}
    I(X,S) =  \sum_{s\in\mathcal{S}} P_S(s) \; \gendivergence{d_\textnormal{KL}}{P_{X\mid S=s}}{P_X}
\label{eq:mi_from_divergence}
\end{equation}

\subsection{Segregation as diversity ratios}\label{sec:mi_from_diversity}
The intuition behind this approach is that when segregation is high, most people live in places that are less diverse than the overall population. Thus, segregation is the `excess' diversity of the population over the average of the spatial units. For a diversity index $H:\Delta_n\to[0,\infty]$, we define
\begin{equation}
    I = H\left(P_X\right) - \sum_{s\in \mathcal{S}} P_S(s) \, H\left( P_{X|S=s} \right).
\end{equation}

This formulation differs from \textcite{Reardon2002}'s only in normalisation. To guarantee that $I$ is non-negative, we would need $H$ to be concave, following the intuition that mixing populations does not decrease diversity. This is \textcite{Reardon2002}'s assumption. How do we choose $H$?

\subsubsection{The chain rule}
To find desirable properties for a diversity index, we turn once again to behaviour under coarse-graining. Let $\tilde{\mathcal{X}}$ denote a set of supergroups of demographic identities and $\tilde{X}$ the corresponding random variable. The diversity at the coarse scale is $H\left(P_{\tilde{X}}\right)$ and within supergroup $\tilde{x}$ is $H\left( P_{X|\tilde{X}=\tilde{x}} \right)$. The chain rule allows us to decompose diversity into between and within-supergroup contributions.

\begin{definition}
    A diversity index $H$ satisfies the \textbf{chain rule} if 
    \begin{equation}
        H\left(P_X\right) = H\left( P_{\tilde{X}} \right) + \sum_{\tilde{x} \in \tilde{\mathcal{X}}} P_{\tilde{X}}(\tilde{x}) \, H\left( P_{X|\tilde{X}=\tilde{x}} \right).
    \end{equation}
\end{definition}

\textcite{Faddeev1956} shows that the only continuous diversity index that satisfies the chain rule is Shannon entropy~\cite{Shannon1948}.
\begin{definition}
    \textbf{Shannon entropy} $\mathcal{H}:\Delta_n \to [0, \log n]$ is given by
    \begin{equation}
        \mathcal{H}(P) = -\sum_{i=1}^n P(i)\, \log P(i),
    \end{equation}
    with the convention $0\log 0=0$.
\end{definition}

We state the variant of Faddeev's characterisation in \textcite{Leinster2021}.
\begin{theorem} (\textcite[Theorem 2.5.1]{Leinster2021})
    If a diversity index $H$ is continuous and satisfies the chain rule, then $H$ is Shannon entropy up to a multiplicative factor.
\end{theorem}
Similar to before, we can assume the factor to be 1 without loss of generality, giving us the second definition of mutual information.
\begin{equation}
     I(X,S) =  \mathcal{H}\left(P_X\right) - \sum_{s\in\mathcal{S}} P_S(s) \, \mathcal{H}\left( P_{X|S=s} \right)
\label{eq:mi_from_entropy}
\end{equation}
The equivalence of \cref{eq:mi_from_divergence,eq:mi_from_entropy} is no coincidence: KL divergence is the Bregman divergence of negative Shannon entropy (see \textcite{Chodrow2025} for details).

\subsection{Segregation as a measure of association}
The idea behind this approach is that in a segregated population, an individual's spatial location provides information about their demographic identity. In a null model with no association, the joint distribution is the product of the marginals $P_X \otimes P_S$ such that the probability of $(x,s)$ is $P_X(x)\, P_S(s)$. Segregation is the divergence of the empirical joint $P_{X,S}$ from $P_X \otimes P_S$. Requiring the mean-minimiser and data processing properties once again gives us KL divergence and the third definition of mutual information.
\begin{equation}
    I(X,S) = \gendivergence{d_{\textnormal{KL}}}{P_{X,S}}{P_X \otimes P_S}
\end{equation}

\textcite{Fullwood2023}'s recent characterisation of mutual information does not assume the functional form of a divergence from the null, providing a more general starting point. In the next section, we interpret the properties of \textcite{Fullwood2023} in the context of segregation. For the properties unique to the ordering of regions by mutual information, see \textcite{Frankel2011}.

\subsubsection{Characterisation of mutual information}

\begin{theorem} (\textcite[Theorem 1]{Fullwood2023})
    Let $F$ be a function that maps pairs of random variables to $[0, \infty]$ and satisfies the following properties.
    \begin{enumerate}[noitemsep,topsep=0pt]
        \item Continuity
        \item Invariance under pullbacks
        \item Vacuity with a constant argument
        \item Symmetry
        \item Strong additivity
        \item Weak functoriality
    \end{enumerate}
    Then $F$ is a non-negative multiple of mutual information. Conversely, mutual information has all these properties.
\end{theorem}
Let us look at what each of the properties mean for segregation.

\paragraph{Continuity.} This is a regularity assumption, ensuring that $I$ changes continuously with $X$ and $S$.

\paragraph{Invariance under pullbacks.} This guarantees that $I$ depends only on the probabilities encoded by $X$ and $S$, not the actual labels.

\paragraph{Vacuity with a constant argument.} $I(X,C)=0$ if $C$ is a constant random variable: segregation is $0$ if there is only one demographic identity or spatial location.

\paragraph{Symmetry.} $I(X,S)=I(S,X)$. Demographic identity and spatial unit are treated mathematically identically. \cref{eq:mi_from_divergence,eq:mi_from_entropy} can be mirrored to write mutual information as follows.
\begin{align}
    I(X,S) &= \sum_{x \in \mathcal{X}} P_X(x)\; \gendivergence{d_{\textnormal{KL}}}{P_{S\mid X=x}}{P_S} \\
    &=  \mathcal{H}(P_S) - \sum_{x \in \mathcal{X}} P_X(x)\, \mathcal{H}(P_{S \mid X=x})
\end{align}
This also makes its decomposition across scales symmetric (see next section).

\paragraph{Strong additivity.} Consider a population separated into a set $\mathcal{Y}$ of `islands' indicated with the random variable $Y$. Each island $y\in\mathcal{Y}$ has its own set of demographic identities $\mathcal{X}^y$ and spatial units $\mathcal{S}^y$. If we coarse-grain demography and space to the island scale -- each island contains identical individuals -- then $Y$ indicates both identity and location and the segregation is $I(Y,Y)$. The segregation within island $y$ is $I(X^y,S^y)$. Strong additivity means that total segregation is the sum of between-island and population-weighted within-island contributions.
\begin{equation}
    I(Y,Y) + \sum_{y \in \mathcal{Y}} P_Y(y) \, I(X^y,S^y)
\end{equation}
This is reminiscent of the chain rule. We note that $I(Y,Y)=\mathcal{H}\left(P_Y\right)$ (see \cref{eq:mi_from_entropy}).

\paragraph{Weak functoriality.} Consider a random variable $Y$ that is constructed using a stochastic transformation $T$ of $X$. Let $Z$ be a third random variable obtained by transforming $Y$ using some $T'$. Intuitively, the information from $X$ to $Z$ goes through $Y$. $(X,Y,Z)$ is a Markov triangle if knowing the outcome of $X$ and $Z$ uniquely determines the outcome of $Y$~\cite{Fullwood2021}. Weak functoriality requires
\begin{equation}
    I(X,Z) = I(X,Y) + I(Y,Z) - I(Y,Y).
\end{equation}
for every Markov triangle $(X,Y,Z)$. In particular, $I(X,Z)\leq I(X,Y) + I(Y,Z)$, which is reminiscent of the data processing property.

Let us say that $X$ indicates ethnicity, $Y$ neighbourhood, and $Z$ school. From the assumptions, neighbourhood is mapped from ethnicity, school is mapped from neighbourhood, and knowing an individual's ethnicity and school determines their neighbourhood. Weak functoriality gives us the value of the segregation of schools by ethnicity. It is the sum of the segregation of neighbourhoods by ethnicity and that of schools by neighbourhood, adjusted by the information in the neighbourhood variable itself, which was counted twice.

Two indicators about an individual uniquely determining a third is unlikely in practice. However, Markov triangles generalise compositions of deterministic maps, and the special case of coarse-graining provides some intuition on the interpretation of weak functoriality. With $\tilde{X}$ indicating supergroups, knowing $X$ and $S$ clearly determines $\tilde{X}$. Thus, $(X,\tilde{X},S)$ is a Markov triangle if the mapping between demographic identities and spatial units is at the supergroup scale. In other words, if $\tilde{X}$ is a sufficient statistic of $X$ for $S$. By weak functoriality,
\begin{equation}
    I(X,S) = I(X,\tilde{X}) + I(\tilde{X},S) - I(\tilde{X},\tilde{X}).
\end{equation}
Because $X$ determines $\tilde{X}$, $I(X, \tilde{X}) = \mathcal{H}( \tilde{X} ) = I( \tilde{X}, \tilde{X} )$. Thus, we have $I(X,S) = I(\tilde{X},S)$. This models the idea that refining demography does not contribute to segregation if it does not reveal more about location.

Mutual information has the bounds $0 \leq I(X,S) \leq \min \left( \mathcal{H}(P_X),\, \mathcal{H}(P_S)\right)$. It is $0$ when all the spatial units are demographically identical and is maximum when either the identity or the spatial unit of an individual uniquely determines the other.

\subsection{The multi-scale decomposition}
\begin{figure}[h!]
    \centering
    \includegraphics{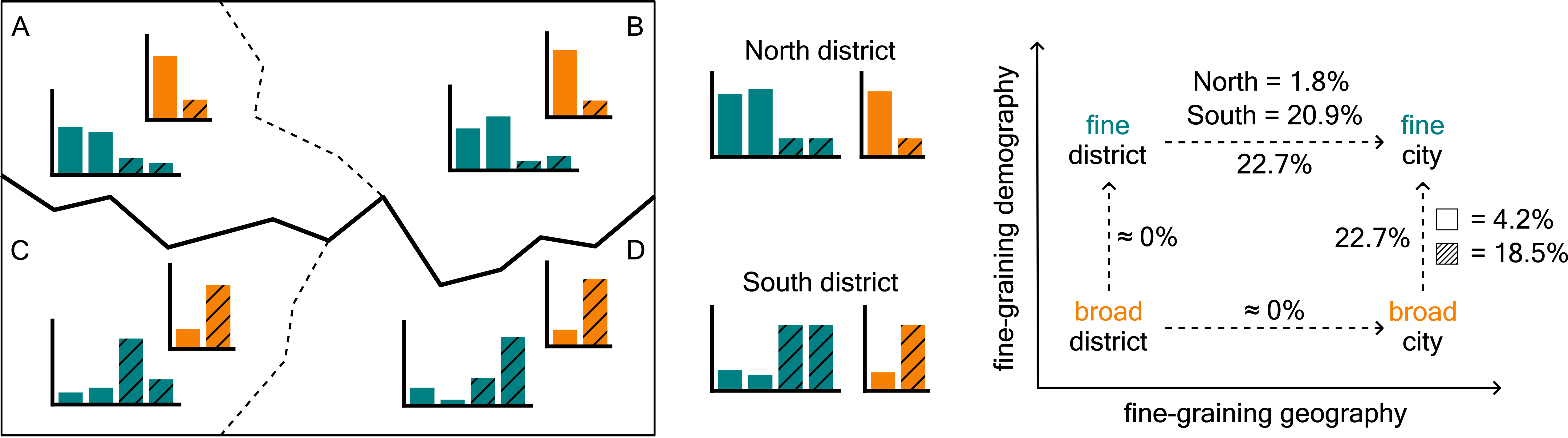}
    \caption{\textbf{Method schematic.} This fictional setting has four \textit{cities} (A-D) aggregated into two \textit{districts}. Cities A and B form the North district and C and D form the South. There are four demographic identities at the \textit{fine} scale (green), aggregated into two \textit{broad} groups (orange; hatched and solid). The bar charts plot the demographic compositions of each city and district. Fine-graining geography from the (fine, district) scale adds 22.7\% of the total information at the (fine, city) scale. Dis-aggregating the North and South districts account for $1.8\%$ and $20.9\%$ respectively, capturing that the South district combines cities with very different demographic compositions. Fine-graining demography from the (broad, city) scale also adds around 22.7\% of the total information. Splitting the hatched and solid groups contribute $18.5\%$ and $4.2\%$ respectively, indicating that the hatched group is more internally segregated. That we lose roughly the same amount of information in both coarse-grainings is peculiar to this simple example and is not generally the case.}
    \label{fig:schematic}
\end{figure}

We illustrate how we can decompose mutual information into contributions at different spatial and demographic scales using a fictional setting (\cref{fig:schematic}). A spatial scale is a partition $\tilde{\mathcal{S}}$ of $\mathcal{S}$, where spatial units are aggregated into regions $\tilde{s}$. In \cref{fig:schematic}, four cities (A-D) are aggregated into two districts (North and South). A demographic scale $\tilde{\mathcal{X}}$ is a coarse-graining of $\mathcal{X}$ into supergroups $\tilde{x}$. \cref{fig:schematic} has four identities coarse-grained into two broad groups (solid and hatched). We denote the corresponding coarse-grained random variables as $\tilde{X}$ and $\tilde{S}$.

Coarse-graining inevitably leads to a loss of information (\cref{sec:dpi}), which decomposes into the contributions of each region and supergroup. At demographic scale $\tilde{\mathcal{X}}$,
\begin{equation}
\begin{split}
    I(\tilde{X}, S) &= I(\tilde{X},\tilde{S}) + \sum_{\tilde{s} \in \tilde{\mathcal{S}}} P_{\tilde{S}}(\tilde{s})\,I^{\tilde{X}}_{\tilde{s}} \textnormal{, where} \\
    I^{\tilde{X}}_{\tilde{s}} &=I\left(\tilde{X}\mid \tilde{S}=\tilde{s}, \, S\mid \tilde{S}=\tilde{s}\right)
\end{split}
\label{eq:spatial_decomposition}
\end{equation}
is the segregation within $\tilde{s}$. The information lost by coarse-graining to $\tilde{s}$ is $P_{\tilde{S}}(\tilde{s})\,I^{\tilde{X}}_{\tilde{s}}$. Cities in the South district have markedly different fine-scale demographic compositions, so the district-level composition is less representative of its cities than in the North (\cref{fig:schematic}). This is identified by the spatial decomposition: $I^\textnormal{fine}_\textnormal{South} \gg I^\textnormal{fine}_\textnormal{North}$.

The symmetry of mutual information permits an identical decomposition into the contributions of supergroups. At spatial scale $\tilde{\mathcal{S}}$,
\begin{equation}
\begin{split}
     I(X, \tilde{S}) &= I(\tilde{X}, \tilde{S}) + \sum_{\tilde{x} \in \tilde{\mathcal{X}}}  P_{\tilde{X}}(\tilde{x})\,I^{\tilde{S}}_{\tilde{x}} \textnormal{, where} \\
    I^{\tilde{\mathcal{S}}}_{\tilde{x}} &= I\left(X\mid \tilde{X}=\tilde{x}, \, \tilde{S}\mid \tilde{X}=\tilde{x}\right)
\end{split}
\label{eq:demographic_decomposition}
\end{equation}
is the segregation within $\tilde{x}$. The hatched group merges identities with very different population shares, particularly in cities C and D (\cref{fig:schematic}) and $I^\textnormal{city}_\textnormal{hatched} \gg I^\textnormal{city}_\textnormal{solid}$.

We assess the quality of a description at scales $( \tilde{\mathcal{X}},\tilde{\mathcal{S}})$ by comparing $I(\tilde{X},\tilde{S})$ to $I(X,S)$, as well as the information captured by fine-graining geography and demography separately: $I(X,\tilde{S})$ and $I(\tilde{X}, S)$. When the scales are defined in a nested hierarchy, we can isolate the incremental contributions of successively finer scales. See Appendix~\ref{sec:glossary} for a glossary of notation and interpretations.

\textcite{Theil1971}'s information theory index is $I(X,S)/\mathcal{H}(P_X)$ and they were the first to use decompositions of this form. However, the normalisation destroys the symmetric decomposability properties that make mutual information uniquely suited for multi-scale analysis~\cite{Mora2011}.

\subsection{Measuring diversity}
The upper bound $I(X,S) \leq \min \left( \mathcal{H}(P_X),\, \mathcal{H}(P_S)\right)$ reflects that segregation is intricately linked to diversity: a homogenous population cannot be segregated. We report diversity using the exponential of Shannon entropy~\cite{Hill1973}.
\begin{definition}
    \textbf{Diversity} $\mathcal{D}:\Delta_n \to [1, n]$ is given by
    \begin{equation}
        \mathcal{D}(P) = \exp \left( \mathcal{H}(P) \right).
    \end{equation}
\end{definition}
$\mathcal{D}$ has the advantage of being an effective number: $\mathcal{D}(P)=k$ means that $P$ is as diverse as the uniform distribution over $k$ groups. However, $\mathcal{D}$ is not concave, preventing us from using it in the diversity ratios approach (\cref{sec:mi_from_diversity}).

\section{Residential ethnic segregation in England and Wales} \label{sec:empirics}

\begin{table}[h!]
    \centering
    \begin{tabular}{lcc}
    \toprule
    \textbf{Ethnic group} & \textbf{Population} & \textbf{Percentage}\\
    \midrule
    \multicolumn{3}{l}{Asian, Asian British or Asian Welsh}\\
    \midrule
         Bangladeshi &  644,912 & 1.08\% \\
         Chinese & 445,528 & 0.75\% \\ 
         Indian & 1,864,300  & 3.13\% \\
         Pakistani & 1,588,040  & 2.66\% \\
         Other Asian & 973,112  & 1.63\% \\
    \midrule
    \multicolumn{3}{l}{Black, Black British, Black Welsh, Caribbean or African}\\
    \midrule
        African  & 1,488,319 & 2.5\% \\
        Caribbean  & 623,054 & 1.05\% \\
        Other Black  & 297,607 & 0.5\% \\
    \midrule
    \multicolumn{3}{l}{Mixed or Multiple ethnic groups}\\
    \midrule
        White and Asian & 488,329 & 0.82\% \\
        White and Black African & 249,689 & 0.42\% \\
        White and Black Caribbean & 513,011 & 0.86\% \\
        Other Mixed or Multiple ethnic groups & 467,310 & 0.78\% \\
    \midrule
    \multicolumn{3}{l}{White}\\
    \midrule
        English, Welsh, Scottish, Northern Irish or British & 44,355,016 & 74.42\% \\
        Irish & 507,399 & 0.85\% \\
        Gypsy or Irish Traveller & 67,644 & 0.11\% \\
        Roma & 101,183 & 0.17\% \\
        Other White & 3,668,273 & 6.15\%\\
    \midrule
    \multicolumn{3}{l}{Other ethnic group}\\
    \midrule
        Arab & 331,774 & 0.56\% \\
        Any other ethnic group & 923,949 & 1.55\% \\
    \bottomrule
    \end{tabular}
    \caption{\textbf{Ethnic composition of England and Wales.} Self-reported ethnic identities of the population of England and Wales in 2021. Source: 2021 Census, Table TS021.}
    \label{tab:ethnic_composition_ew}
\end{table}
\begin{table}[h!]
    \centering
    \begin{tabular}{lcccc}
    \toprule
    \textbf{Scale} & \textbf{\# units} & \multicolumn{3}{c}{\textbf{Population (percentile)}}\\
    & & 25th & 50th & 75th\\
    \midrule
    OA & 188,880 & 263 & 306 & 355 \\
    LSOA & 35,672  & 1,439 & 1,605 & 1,832 \\
    MSOA & 7,264 & 6,823 & 7,957 & 9,292 \\
    LAD & 331 & 103,295 & 141,928 & 217,786 \\
    \bottomrule
    \end{tabular}
    \caption{\textbf{Spatial scales of England and Wales.} Statistics on the number of units and population of the 25th, 50th, and 75th percentile for different reporting units. OA: Output Area, LSOA: Lower layer Super Output Area, MSOA: Middle layer Super Output Area, LAD: Local Authority District. Source: 2021 Census, Open Geography Portal, ONS.}
    \label{tab:geography_hierarchy_ew}
\end{table}

We demonstrate the ability of decompositions to identify patterns across scales by analysing residential ethnic segregation in England and Wales. Ethnicity in the 2021 census of England and Wales is organised into two levels. There are five ethnic groups at the \textit{broad} scale, of which the \textit{White} group is the majority (82\%) and the \textit{Asian, Asian British or Asian Welsh} (hereafter \textit{Asian}) is the largest minority (9\%). The diversity of England and Wales on the broad demographic scale is $\mathcal{D}(P_\textnormal{broad}) = 2.0$. The supergroups disaggregate into 19 \textit{fine} ethnic identities, of which \textit{White: English, Welsh, Scottish, Northern Irish or British} (hereafter \textit{White British}) are the majority (74\%) and \textit{Other White} are the largest minority (6\%). The largest non-\textit{White} minorities are \textit{Indian} and \textit{Pakistani} (both around 3\%). $\mathcal{D}(P_\textnormal{fine}) = 3.4$, reflecting the low populations of most minority ethnicities (\cref{tab:ethnic_composition_ew}).

The census employs a hierarchy of spatial scales. The finest units are Output Areas (OAs), which nest successively into Lower layer Super Output Areas (LSOAs), Middle layer Super Output Areas (MSOAs), and Local Authority Districts (LADs or districts). The population size varies from a median of around 300 per OA to around 140,000 in a district (\cref{tab:geography_hierarchy_ew}).

\subsection{Segregation at different scales}

\begin{figure}
    \centering
    \includegraphics{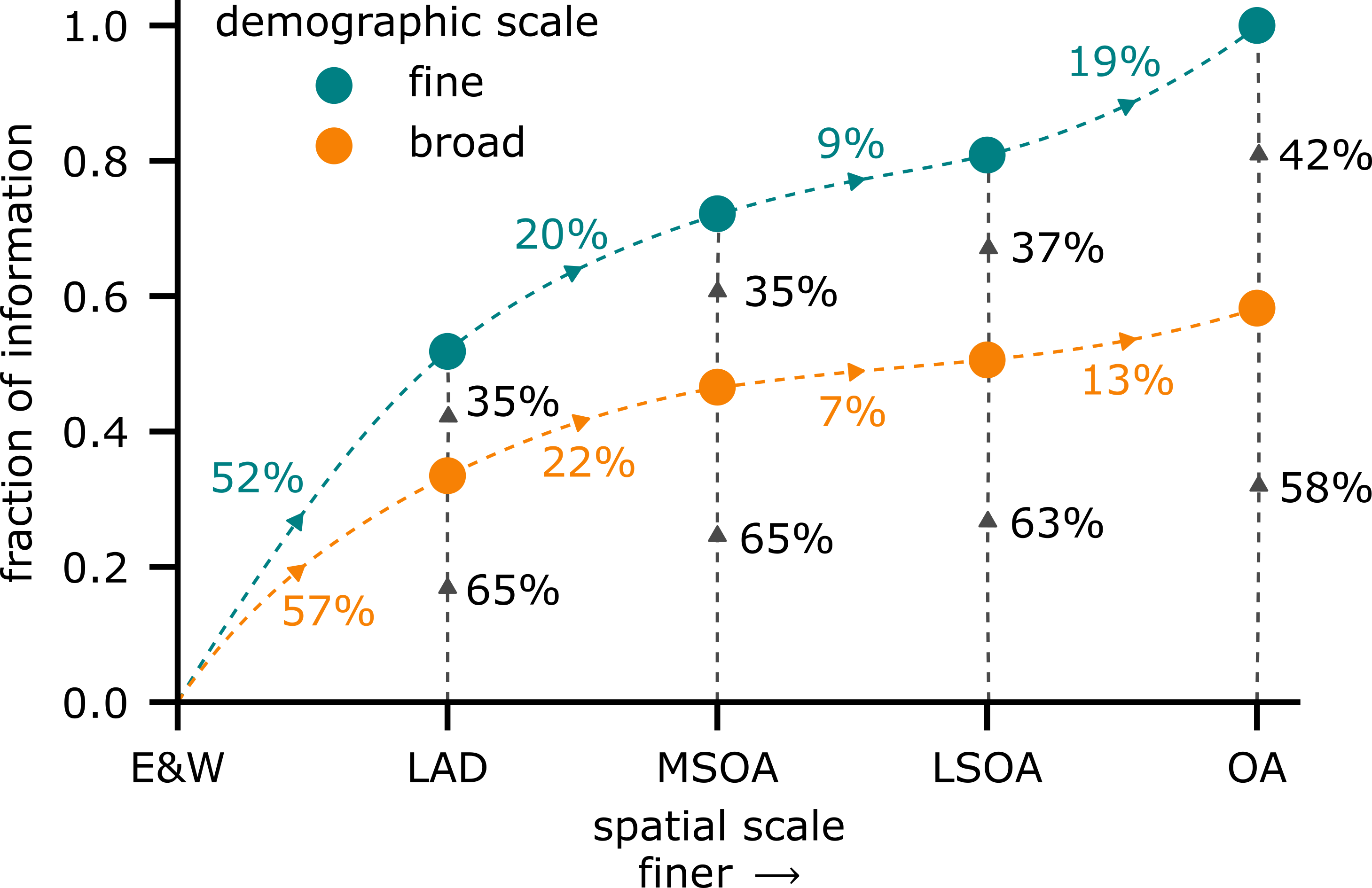}
    \caption{\textbf{Segregation at different scales in England and Wales.} Each marker denotes a choice of spatial (x-axis) and demographic (colour) scale. We plot the amount of information captured at the scale (y-axis) as a fraction of the maximum: $I(\textnormal{fine}, \textnormal{OA})$. For demographic scale $\tilde{\mathcal{X}}$, we can follow spatial fine-graining (left to right) from the coarsest scale with a single region (E\&W) to the finest (OA). We annotate the dashed lines with the information added as a percentage of $I(\tilde{X},\textnormal{OA})$. For instance, going from MSOAs to LSOAs adds $7\%$ of the information at the broad demographic scale. Similarly, for spatial scale $\tilde{\mathcal{S}}$, we annotate demographic fine-graining (bottom to top) with the information added as a percentage of $I(\textnormal{fine},\tilde{S})$. The ability to track the incremental contributions of nested scales is a product of mutual information's multi-scale decomposition.}
\label{fig:multiscale_decomposition}
\end{figure}

How much segregation is captured at different spatial and demographic scales? We plot $I(\tilde{X},\tilde{S})$ as a fraction of the total segregation $I(\textnormal{fine}, \textnormal{OA})$ in \cref{fig:multiscale_decomposition}. The trends of information gained by fine-graining spatially are similar for both demographic scales. More than half the segregation is explained by inter-district differences, suggesting that demography varies widely across districts. Going from MSOAs to LSOAs provides the least additional information: the LSOAs within an MSOA generally have similar demographic compositions. The sharp increase to the OA scale indicates that LSOAs tend to be internally segregated.

While the impact of demographic scale varies by region (\cref{fig:hist_b2f_ratio}), its total effect is similar at the coarser spatial scales: around 65\% of the segregation is between broad ethnic groups. At the OA scale, this goes down to 58\%, suggesting that finer ethnic differences are more important.

In the following sections, we investigate what the spatial and demographic decompositions reveal about segregation at the district scale. We pick districts for this study since they are large enough to be interpretable but still capture a majority of the segregation. The framework applies identically at other scales.

\subsection{Diversity and segregation at the district scale}
The 331 districts vary widely in diversity (\cref{fig:dist_diversity_lads}). Predominantly \textit{White British} districts such as Allerdale and Copeland have diversities as low as $1.25$ on the fine demographic scale. London is the most diverse region. The average diversity of its 33 districts is $8.1$, with the boroughs of Brent and Newham exceeding $11$. While more populous districts are likely to be more diverse\footnote{Pearson $r$ 0.41 (fine) and 0.45 (broad demographic scale), both with p-values $\ll 10^{-12}$.}, 75\% of the districts -- accounting for 64\% of the population -- are less diverse than the overall demographic composition (\cref{fig:line_cumpop_vs_diversity_lads}).

For either demographic scale, we decompose $I(\tilde{X}, \textnormal{OA})$ into $I(\tilde{X}, \textnormal{district})$ and the information within districts. The divergence of the ethnic composition of district $\tilde{s}$ from that of the population is a measure of how atypical it is. Scaled by population, this is also its contribution to $I(\tilde{X}, \textnormal{district})$. The most and least diverse districts are the most atypical (\cref{fig:cuminfexplained_vs_diversity}).

\begin{figure}
    \centering
    \includegraphics{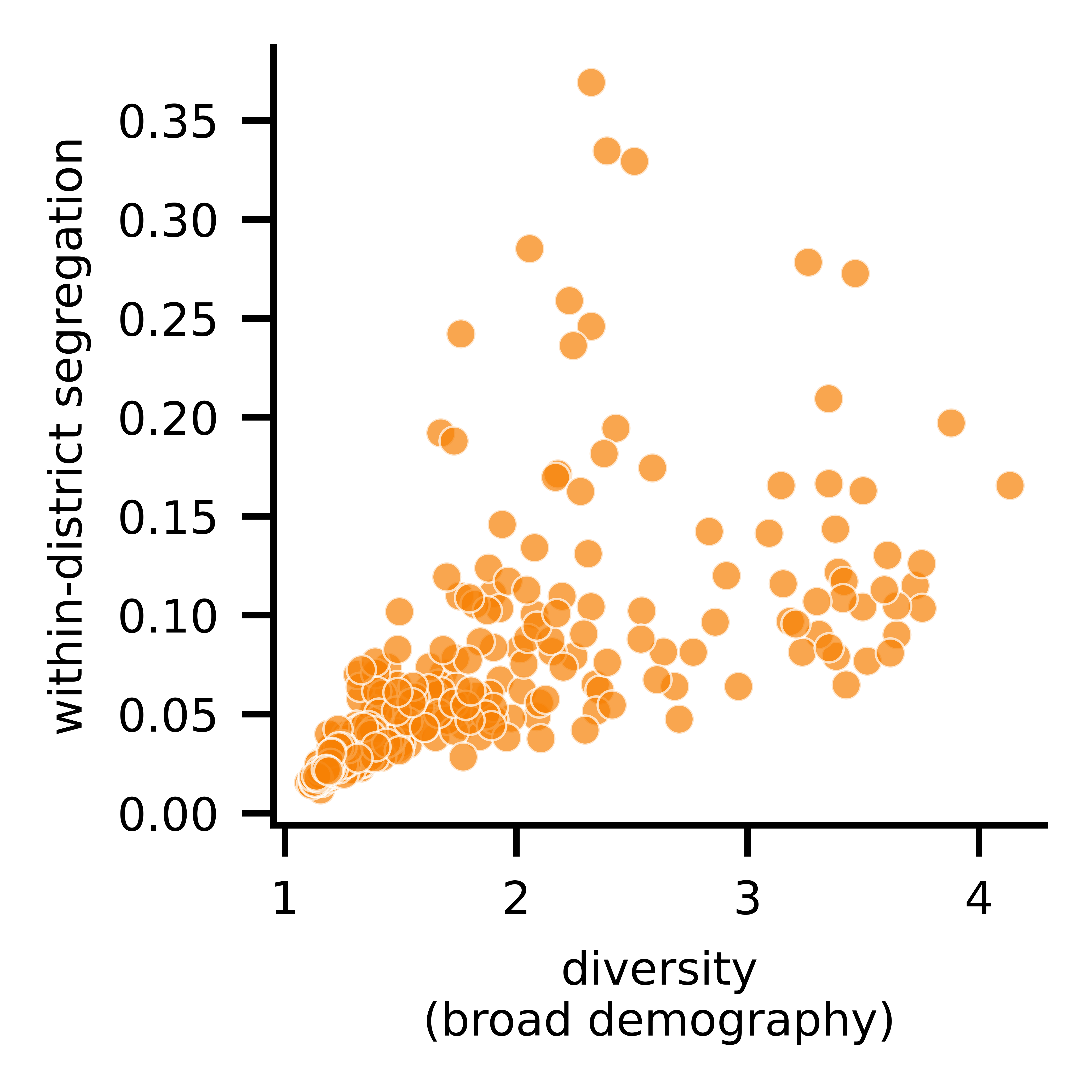}
    \caption{\textbf{Within-district segregation against diversity.} The within-district segregation $I_{\tilde{s}}^{\textnormal{broad}}$ (y-axis) against diversity (x-axis) at the broad demographic scale. Each point corresponds to a district. The wide range of segregation for relatively diverse districts shows that diversity does not necessarily lead to segregation. The plot is qualitatively identical at the fine demographic scale (\cref{fig:scatter_segregation_vs_diversity_fine_lads}).}
\label{fig:scatter_segregation_vs_diversity_broad_lads}
\end{figure}

\begin{figure}
    \centering
    \includegraphics{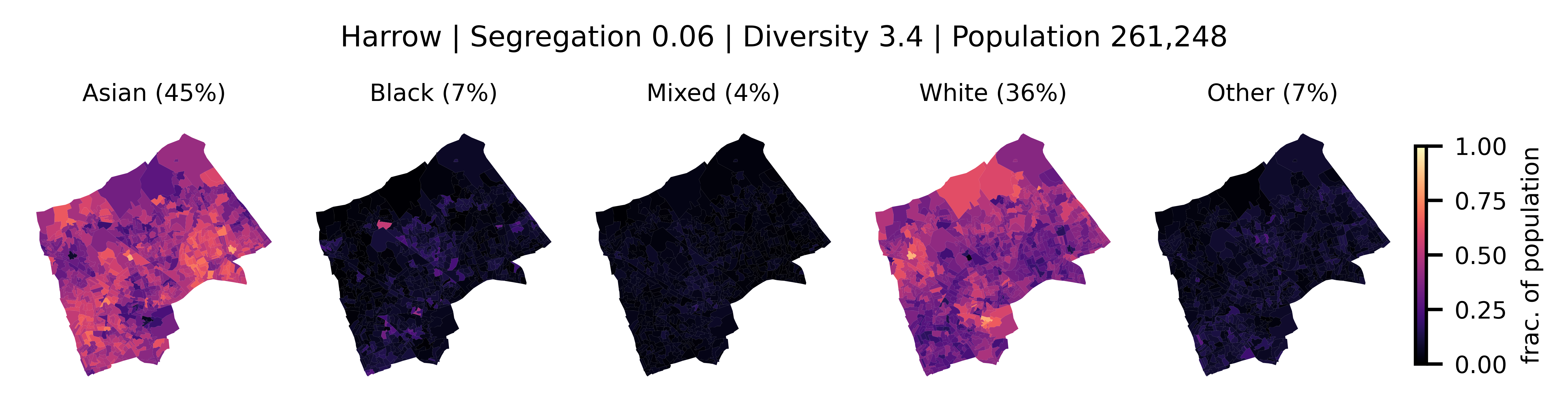}\\
    \includegraphics{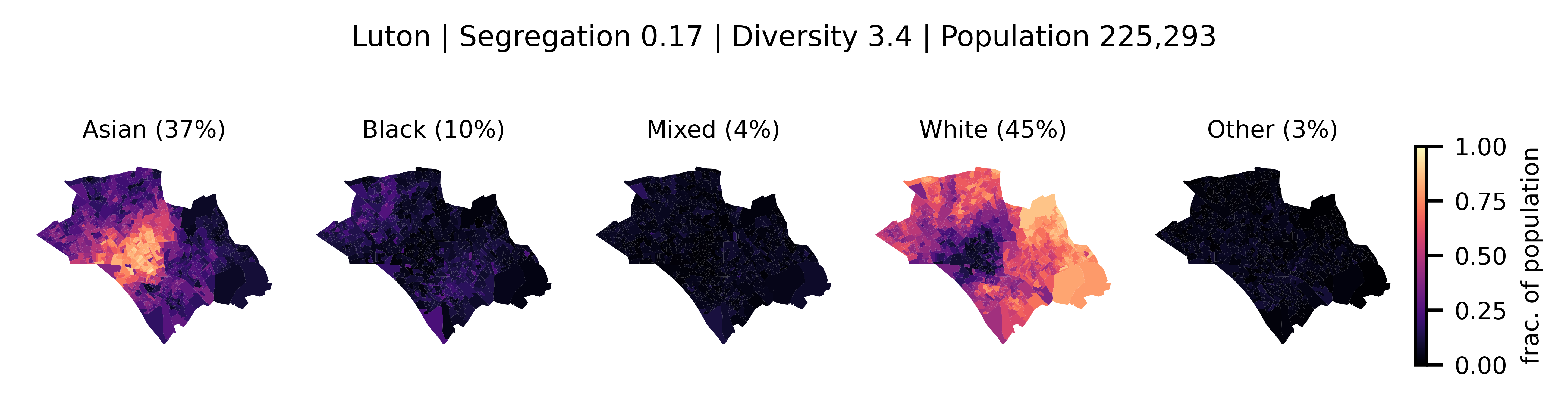} 
    \caption{\textbf{Spatial distribution of ethnic groups in Harrow and Luton.} Each subplot depicts the distribution of a broad ethnic group within Harrow (top) and Luton (bottom) at the OA scale. The colour of an OA indicates the fraction of its population that belongs to the ethnic group. While they have similar demographic compositions, Harrow and Luton show very different levels of segregation.}
\label{fig:maps_harrow_luton}
\end{figure}

How segregated are the districts? There is a positive correlation\footnote{Pearson $r$ 0.66 (fine) and 0.65 (broad demographic scale), both with p-values $\ll 10^{-16}$.} between $I_{\tilde{s}}^{\tilde{\mathcal{X}}}$ and diversity with more variability at higher diversities (\cref{fig:scatter_segregation_vs_diversity_broad_lads}). The correlation is expected: districts without diversity cannot be segregated. However, the wide range of $I_{\tilde{s}}^{\tilde{\mathcal{X}}}$ at high diversities shows that equally diverse districts can be segregated to very different extents. For instance, Harrow and Luton have comparable diversity and population sizes but very different spatial organisation (\cref{fig:maps_harrow_luton}). Both districts have large \textit{White} and \textit{Asian} populations. In Harrow, these groups are distributed relatively evenly ($I_{\textnormal{Harrow}}^{\textnormal{broad}}=0.06$), whereas Luton exhibits pronounced segregation with \textit{White} neighbourhoods surrounding an \textit{Asian} core ($I_{\textnormal{Luton}}^{\textnormal{broad}}=0.17$). We also find a similar contrast between Barnet and Leicester (\cref{fig:maps_barnet_leicester}).

Controlling for diversity is not straightforward. The theoretical maximum of segregation occurs only in the extreme case where each spatial unit contains a single ethnic group. Thus, normalising by the maximum is not particularly informative. Null models of population distribution would provide a more meaningful baseline (see Discussion).

\subsection{Demographic decomposition of segregation in Croydon}
\begin{table}[h!]
    \centering
    \begin{tabular}{cccccc}
        \toprule
        \makecell{\textbf{Broad}\\{\textbf{group}}} & \textbf{Population} & \textbf{Breakdown} & \makecell{\textbf{Within-group}\\{\textbf{diversity}}} & \makecell{\textbf{Within-group}\\{\textbf{segregation}}} & \makecell{\textbf{Information}\\{\textbf{lost}}}\\ \midrule
        Asian & 18\% & \makecell{8\% Indian\\ 4\% Pakistani\\ 4\% Other} & 3.9 & 0.19 & 32\% \\
        \midrule
        Black & 23\% & \makecell{10\% African \\ 9\% Caribbean \\ 3\% Other} & 2.7 & 0.06 & 12\%\\
        \midrule
        White & 48\% & \makecell{37\% British \\ 9\% Other} & 1.9 & 0.09 & 39\% \\
        \bottomrule
    \end{tabular}
    \caption{\textbf{Demography of Croydon.} The largest broad ethnic groups in Croydon along with the population shares of their largest fine-scale ethnicities. The within-group diversity is the effective number of fine ethnicities within $\tilde{x}$. Segregation is the population-normalised within-supergroup value and information lost is as a fraction of the total loss by coarse-graining demography.}
    \label{tab:croydon}
\end{table}

\begin{figure}[h!]
    \centering
    \includegraphics{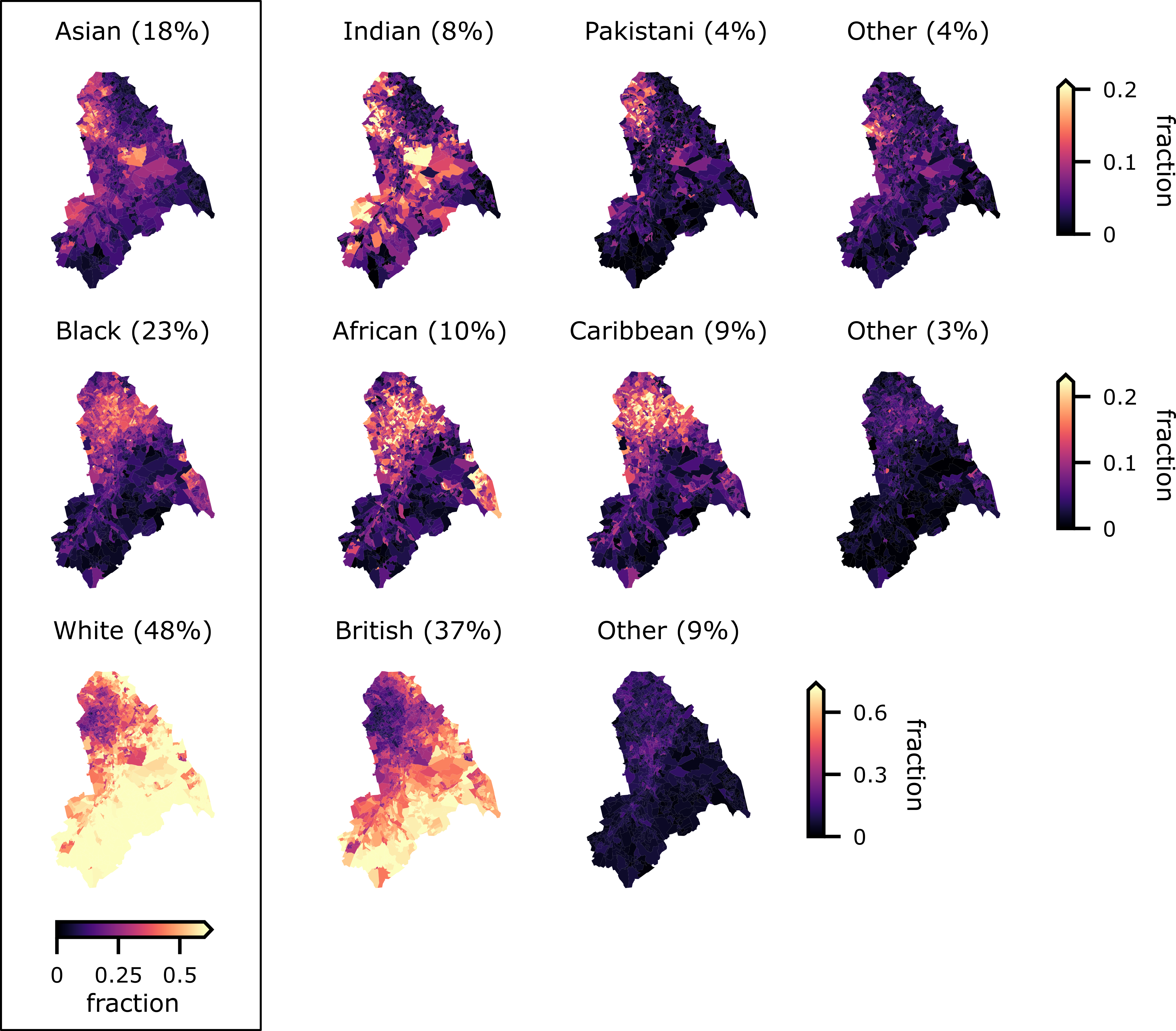}
    \caption{\textbf{Spatial distribution of ethnic groups in Croydon.} Each subplot depicts the distribution of an ethnic group at the OA scale. The colour of an OA indicates the fraction of its population that belongs to the ethnic group. We annotate each subplot with population share. The box on the left has the distribution of the broad \textit{Asian}, \textit{Black}, and \textit{White} groups, with the colour scale at the bottom. Each row depicts the distribution of the largest fine-scale ethnicities of the corresponding group with colour scales on the right. We truncate colour scales to make variation in the distributions of the minorities visible.}
    \label{fig:croydon}
\end{figure}
The London borough of Croydon is one of the most diverse districts in England and Wales. Croydon has large \textit{Asian}, \textit{Black}, and \textit{White} populations, with a broad scale diversity of $3.7$ (ranked $5^\textnormal{th}$) and a fine-scale diversity of $9.1$ (ranked $8^\textnormal{th}$). Croydon is the $40^\textnormal{th}$ most segregated district (measured by $I_{\tilde{s}}^{\tilde{X}}$) on both demographic scales and has clearly segregated neighbourhoods (\cref{fig:croydon}).

How much within-Croydon segregation is lost by coarse-graining demographic identities? $1 - I^\textnormal{broad}_\textnormal{Croydon} / I^\textnormal{fine}_\textnormal{Croydon} = 48\%$. We decompose this into the contributions of each ethnic supergroup (\cref{eq:demographic_decomposition}), showing that they are internally segregated to different extents (\cref{tab:croydon}). Despite having a smaller share of the population, \textit{Asians} account for 32\% of the within-group segregation, while the \textit{Black} supergroup contributes 12\%. This difference is apparent in their spatial distribution (\cref{fig:croydon}). Both major \textit{Black} ethnicities, \textit{African} and \textit{Caribbean}, live mainly in the north of Croydon, with some \textit{Africans} in the east. The \textit{Asian} ethnicities have less overlap. \textit{Pakistanis} and \textit{Other Asians} live primarily in the north west, while \textit{Indians} also reside in the centre and the south west. Within-\textit{White} segregation is also strong, with \textit{White British} in the south and \textit{Other White} in the north with the non-\textit{Whites}.

The latter observation sparks the question: Are \textit{White} minorities more likely to live amongst non-\textit{Whites} than \textit{British Whites}? To investigate this, we compare two different aggregations of identities into two supergroups. The \textit{White} / non-\textit{White} grouping captures 30\% of the within-Croydon information, while the \textit{White British} / non-\textit{White British} retains 37\%, showing that it is a better explanation of the empirical segregation. This pattern is not universal: 21 London boroughs display the opposite relationship (\cref{fig:scatter_2group_coarsegraining_comparison,fig:maps_croydon_ealing}), suggesting that distinct social processes may be at work. That the effectiveness of demographic scales is region-dependent suggests that developing and comparing bespoke demographic coarse-grainings is a means to understanding local social structure.

\section{Discussion}

In this work, we derived mutual information from each of \textcite{Reardon2002}'s general approaches for constructing a multi-group segregation index. In their disproportionality-based approach, segregation is the demographic dissimilarity of a randomly-sampled individual's spatial unit from the population. Operationalising this involves picking a divergence to compare demographic compositions. We identified the mean-minimiser and data processing properties as desirable for multi-scale analyses. Only KL divergence satisfies them both simultaneously~\cite{Jiao2014} and mutual information is the corresponding index. In the diversity-based approach, segregation is the difference between the diversity of the population and the spatial unit of a randomly-sampled individual. To implement this, we need to pick a diversity index. Requiring a decomposition into between and within-supergroup diversities narrows the choice down to Shannon entropy~\cite{Faddeev1956}, leading once again to mutual information. In the association-based approach, segregation is the association between the demographic identity and spatial unit of an individual. \textcite{Reardon2002}'s functional form involves choosing a divergence, which has to be KL divergence if mean-minimiser and data processing are desired. We discussed the properties in \textcite{Fullwood2023}'s characterisation of mutual information as a more general starting point.

Together, these insights suggest that mutual information's decomposability is quite special~\cite{Mora2011, Frankel2011}. We demonstrated its usefulness with an exploratory analysis of residential ethnic segregation in England and Wales. Mutual information distinguishes between regions such as Luton and Harrow with similar diversity but very different levels of segregation. Within Croydon, we identified \textit{Asians} and \textit{Whites} as being more internally segregated than the \textit{Black} supergroup. A more detailed multi-scale study and extensions to other regions, demographic indicators, and non-residential contexts are natural directions for future research.

Our results suggest that the choice of spatial and demographic scales depend on each other. The information captured can serve as an objective function to construct bespoke scales from data~\cite{Chodrow2017}. An informative spatial scale is one with homogeneous regions whose boundaries capture demographic transitions. Constructing one can be tackled with a wide array of clustering tools, including spatially constrained ones~\cite{Chodrow2017}. Maximising the information of a demographic scale corresponds to grouping demographic identities such that individuals within a supergroup are well-mixed. When dealing with multi-dimensional demography data, this can be used to identify which attributes are most closely associated with segregation~\cite{Fischer2003, Iceland2006}.

The main limitation of our analysis is that it does not assess statistical significance. Comparing empirical estimates against suitable null models is an important next step, linking this framework to multilevel segregation modelling~\cite{Leckie2012, Jones2015}. A suite of null models would enable hypothesis testing and statistical inference of informative scales. On a more theoretical level, mutual information has the same limitation as every other index: it encodes a single notion of segregation. For instance, in the diversity-based approach, the contribution of an individual of demographic group $x$ to Shannon entropy is $\log (1/P_X(x))$. While this functional form uniquely satisfies the chain rule, it is a specific choice of how to weigh the influence of groups. Ecologists have argued for the use of diversity indices with an interpretable parameter that controls the relative importance of abundant and rare species~\cite{Hill1973}. The same strategy would allow researchers to tune the contributions of majority and minority groups to the segregation index, leading to more thorough comparisons.

\section{Methods}

\subsection{Data description: Census of England and Wales}

The census of England and Wales is conducted decennially by the Office for National Statistics (ONS) and collects demographic information from individuals and households. We obtain ethnicity data from Table TS021 of the 2021 census via Nomis~\cite{Nomis}. The census sampled 59.6 million individuals, who were asked to select one of 19 ethnic groups that ‘best describe your ethnic group or background’ (see \cref{tab:ethnic_composition_ew}). The reporting areas form a nested hierarchy of spatial scales (see \cref{tab:geography_hierarchy_ew}). The units at the finest scale, Output Areas (OAs) are designed to be homogeneous in housing tenure and dwelling type~\cite{Martin2001}. We obtain the geographies of the OAs from the ONS~\cite{UKShapefiles} and the hierarchy of reporting areas from the Open Geography Portal~\cite{UKGeographyLookup}.

\subsection{Software}
We implement our methods in Python 3.12 using standard libraries including NumPy~\cite{NumPy}, pandas~\cite{Pandas}, and GeoPandas~\cite{Geopandas} for analysis and matplotlib~\cite{Matplotlib} and seaborn~\cite{Seaborn} for visualisations. Our software and analysis is public (see our code release~\cite{GitHub}).

\section{Declarations}
\paragraph{Ethics approval and consent to participate.} Not applicable.

\paragraph{Consent for publication.} Not applicable.

\paragraph{Availability of data and materials.} The data analysed in this study are made public by the ONS~\cite{Nomis, UKShapefiles, UKGeographyLookup}. We include the data in the GitHub repository that accompanies this paper~\cite{GitHub}.

\paragraph{Competing interests.} We declare no competing interests.

\paragraph{Funding.} R.S. is funded by the Mathematical Institute at the University of Oxford. R.L. acknowledges funding from EPSRC grants EP/V013068/1, EP/V03474X/1, and EP/Y028872/1.

\paragraph{Authors' contributions.} R.S.: Conceptualisation, Methodology, Investigation, Visualisation, Writing—original draft, and Writing—review and editing. R.L.: Conceptualisation, Methodology, Writing—review and editing, and Supervision.

\paragraph{Acknowledgements.} We are grateful to Karel Devriendt for many helpful discussions. We thank the ONS for maintaining public data.

\printbibliography

\newpage
\appendix
\setcounter{figure}{0}
\renewcommand{\figurename}{Fig.}
\renewcommand{\thefigure}{S\arabic{figure}}

\begin{center}
    {\Huge Appendix}
\end{center}

\section{Glossary of terms and interpretations}\label{sec:glossary}
\subsection{Scales}
\begin{description}
    \item[$\mathcal{X}$] The set of demographic labels at the finest scale.
    \item[$\mathcal{S}$] The set of spatial units at the finest scale.
    \item[$X$] The random variable that takes values in $\mathcal{X}$ for every individual.
    \item[$S$] The random variable that takes values in $\mathcal{S}$ for every individual.
    \item[$x$] A generic element of $\mathcal{X}$ -- a demographic label.
    \item[$s$] A generic element of $\mathcal{S}$ -- a spatial unit.
    \item[$\tilde{\mathcal{X}}$] A demographic scale -- a partition of $\mathcal{X}$ into broad groups.
    \item[$\tilde{\mathcal{S}}$] A spatial scale -- a partition of $\mathcal{S}$ into regions.
    \item[$\tilde{X}$] The random variable that takes values in $\tilde{\mathcal{X}}$ for every individual.
    \item[$\tilde{S}$] The random variable that takes values in $\tilde{\mathcal{S}}$ for every individual.
    \item[$\tilde{x}$] A generic element of $\tilde{\mathcal{X}}$ -- a broad demographic label.
    \item[$\tilde{s}$] A generic element of $\tilde{\mathcal{S}}$ -- a region.
\end{description}

\subsection{Probability distributions}
\begin{description}
    \item[$P_{\tilde{X},\tilde{S}}$] The joint distribution of demographic identity and spatial unit at the scales $\left(  \tilde{\mathcal{X}}, \tilde{\mathcal{S}} \right)$. $P_{\tilde{X},\tilde{S}}(\tilde{x}, \tilde{s})$ is the probability that a randomly sampled individual belongs to demographic group $\tilde{x}$ and region $\tilde{s}$.
    
    \item[$P_{\tilde{X}}$] The demographic composition of the population at the scale $\tilde{\mathcal{X}}$. 
    \[ P_{\tilde{X}}(\tilde{x}) = \sum_{\tilde{s} \in \tilde{\mathcal{S}}} P_{\tilde{X},\tilde{S}}(\tilde{x}, \tilde{s}) \]
    
    \item[$P_{\tilde{S}}$] The geographic distribution of the population at the scale $\tilde{\mathcal{S}}$. 
    \[ P_{\tilde{S}}(\tilde{s}) = \sum_{\tilde{x} \in \tilde{\mathcal{X}}} P_{\tilde{X},\tilde{S}}(\tilde{x}, \tilde{s}) \]
    
    \item[$P_{\tilde{X}|\tilde{S}=\tilde{s}}$] The demographic composition of region $\tilde{s}$. 
    \[P_{\tilde{X}|\tilde{S}=\tilde{s}}\left( \tilde{x} \right) = P_{\tilde{X},\tilde{S}}\left( \tilde{x},\tilde{s} \right) / P_{\tilde{S}}(\tilde{s})\]
    
    \item[$P_{X|\tilde{X}=\tilde{x}}$] The population distribution over fine-scale identities $x$ within $\tilde{x}$.
    \[P_{X|\tilde{X} = \tilde{x}}\left( x \right) = \begin{cases}
        \frac{P_X(x)}{P_{\tilde{X}}(\tilde{x})} & \textnormal{if } x \in \tilde{x} \\
        0 & \textnormal{otherwise}
    \end{cases}
    \]
    
    \item[$P_{X|\tilde{X}=\tilde{x},\tilde{S}=\tilde{s}}$] The population distribution over fine-scale identities $x$ within $\tilde{x}$ in region $\tilde{s}$. 
    \[P_{X|\tilde{X}=\tilde{x},\tilde{S}=\tilde{s}}(x) = \begin{cases}
        \frac{P_{X|\tilde{S}=\tilde{s}}(x) }{ P_{\tilde{X}|\tilde{S}=\tilde{s}}(\tilde{x})} & \textnormal{if } x \in \tilde{x} \\
        0 & \textnormal{otherwise}
    \end{cases}
    \]

    \item[$P_X\otimes P_S$] The joint distribution at scales $\left(  {\mathcal{X}}, {\mathcal{S}} \right)$ under the null model assuming no segregation.
    \[
        P_X\otimes P_S(x,s) = P_X(x) \times P_S(s)
    \]
\end{description}

\subsection{Information quantities and interpretations}
\begin{description}
    \item[$I(\tilde{X}, \tilde{S})$] The information / segregation at scales $(\tilde{\mathcal{X}}, \tilde{\mathcal{S}})$. \textit{How much segregation is captured by a representation at the given scales?}

    \item[$I(\tilde{X}, \tilde{S}) \,/ \, I(X, S)$] \textit{What fraction of information is retained by coarse-graining demographically and spatially to $(\tilde{\mathcal{X}}, \tilde{\mathcal{S}} )$?}

     \item[$I(\tilde{X}, \tilde{S}) \,/ \, I(X, \tilde{S})$] \textit{What fraction of information is retained by coarse-graining demographically to scale $\tilde{\mathcal{X}}$ at spatial scale $\tilde{\mathcal{S}}$?}

     \item[$I(\tilde{X}, \tilde{S}) \,/ \, I(\tilde{X}, S)$] \textit{What fraction of information is retained by coarse-graining spatially to scale $\tilde{\mathcal{S}}$ at demographic scale $\tilde{\mathcal{X}}$?}

     \item[$I^{\tilde{\mathcal{X}}}_{\tilde{s}}$] \textit{How segregated is region $\tilde{s}$ at demographic scale $\tilde{\mathcal{X}}$?} This is a population-normalised measure that treats regions equally.
     \[ 
        I^{\tilde{\mathcal{X}}}_{\tilde{s}} = I \left(\tilde{X}\mid \tilde{S}=\tilde{s}, \, S\mid\tilde{S}=\tilde{s} \right)
     \]

     \item[$P_{\tilde S}(\tilde{s}) I^{\tilde{\mathcal{X}}}_{\tilde{s}}$] \textit{How much information is lost by coarse-graining spatially to region $\tilde{s}$ at demographic scale $\tilde{\mathcal{X}}$?} Alternatively, \textit{How much information is contained within $\tilde{s}$ at demographic scale $\tilde{\mathcal{X}}$?}

     \item[$I^{\tilde{\mathcal{S}}}_{\tilde{x}}$] \textit{How segregated is supergroup $\tilde{x}$ at spatial scale $\tilde{\mathcal{S}}$?} This is a population-normalised measure that treats supergroups equally.
     \[ 
        I^{\tilde{\mathcal{S}}}_{\tilde{x}} = I \left(X\mid \tilde{X}=\tilde{x}, \, \tilde{S}\mid\tilde{X}=\tilde{x} \right)
     \]

     \item[$P_{\tilde X}(\tilde{x}) I^{\tilde{\mathcal{S}}}_{\tilde{x}}$] \textit{How much information is lost by coarse-graining demographically to group $\tilde{x}$ at spatial scale $\tilde{\mathcal{S}}$?} Alternatively, \textit{How much information is contained within $\tilde{x}$ at spatial scale $\tilde{\mathcal{S}}$}

\end{description}

\newpage
\section{Supplementary Figures}

\begin{figure}[h]
    \centering
    \includegraphics{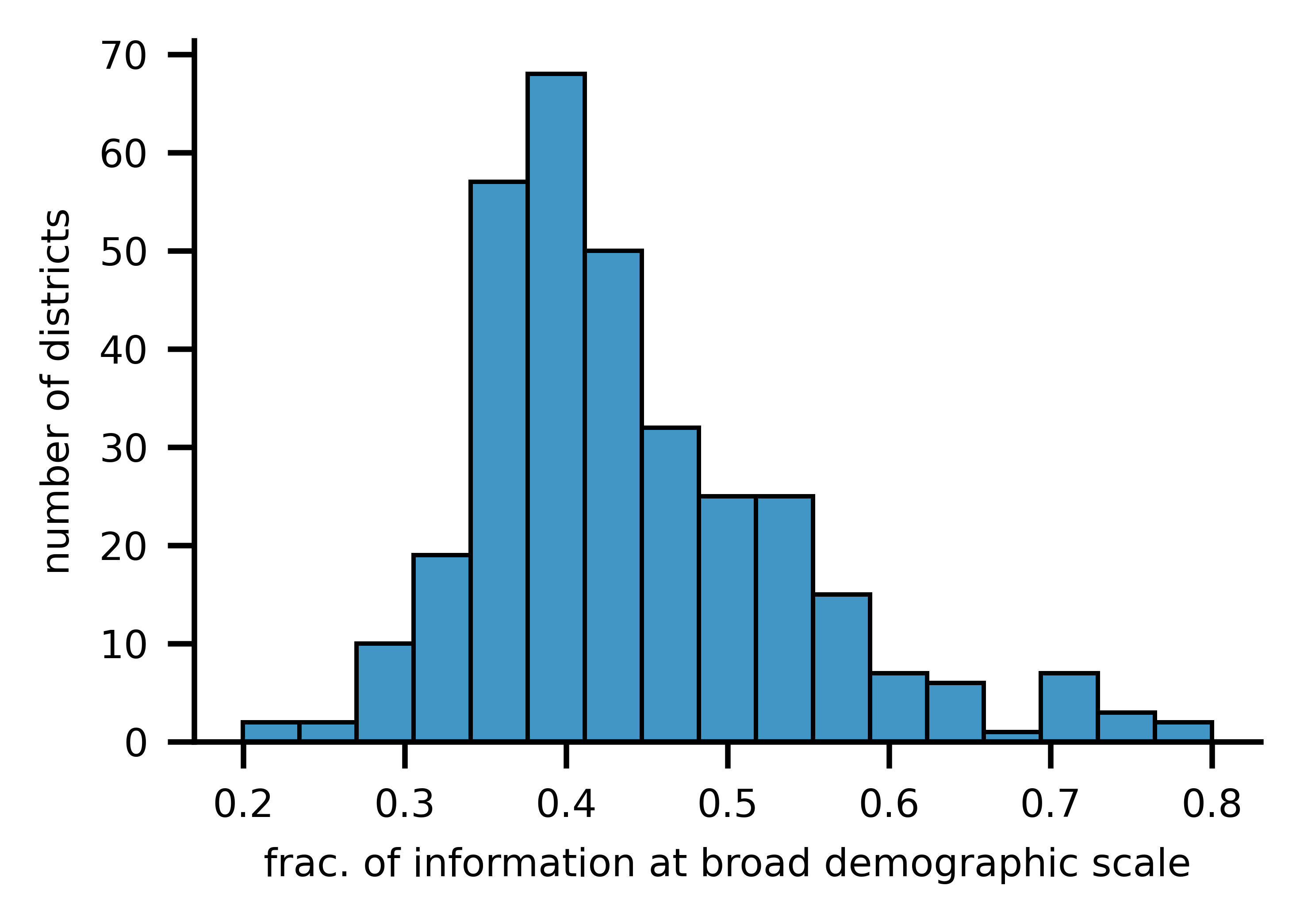}
    \includegraphics{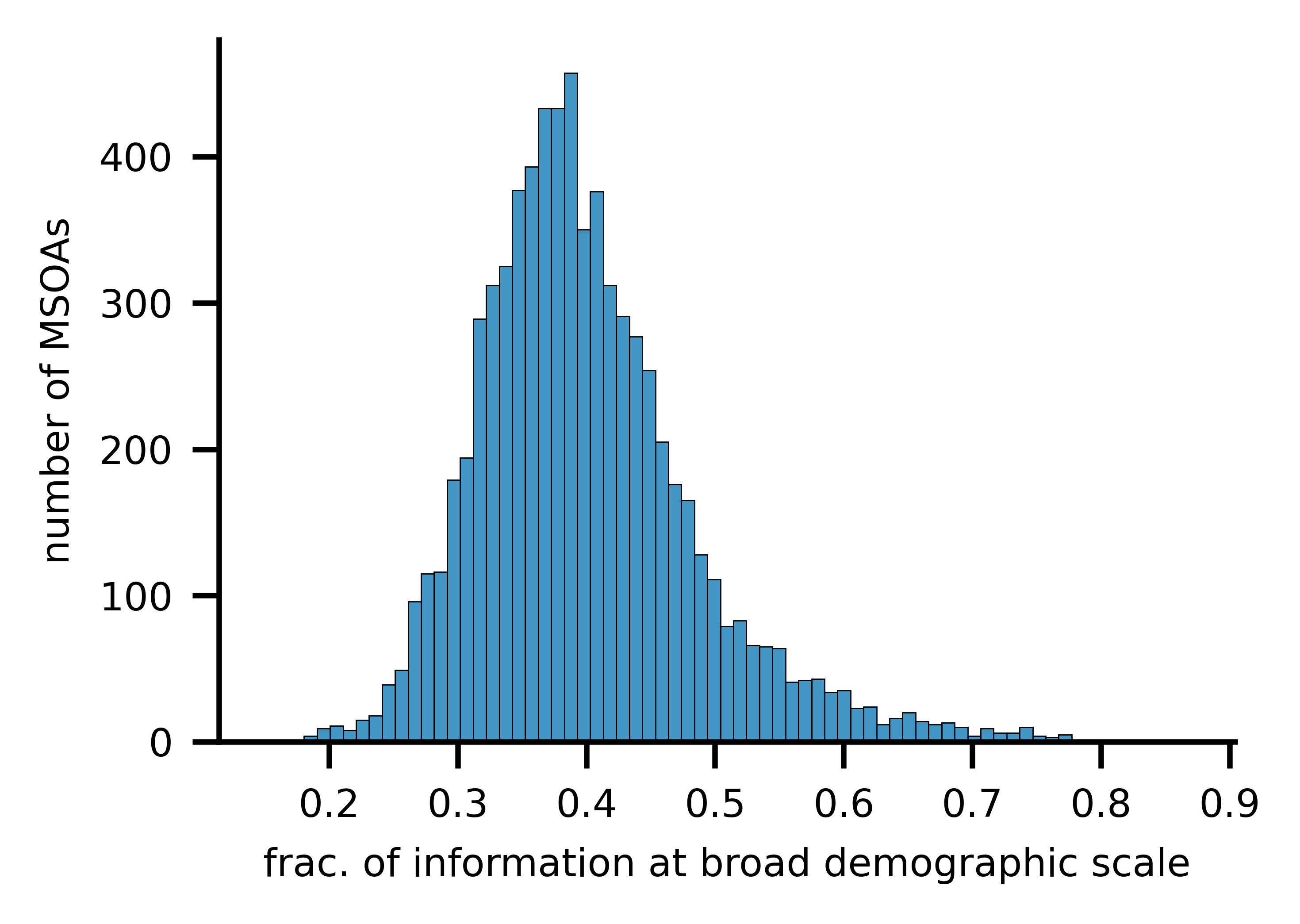}
    \caption{\textbf{The fraction of information captured at the broad demographic scale.} We plot the distribution for districts (left) and MSOAs (right).}
\label{fig:hist_b2f_ratio}
\end{figure}

\begin{figure}[h]
    \centering
    \includegraphics{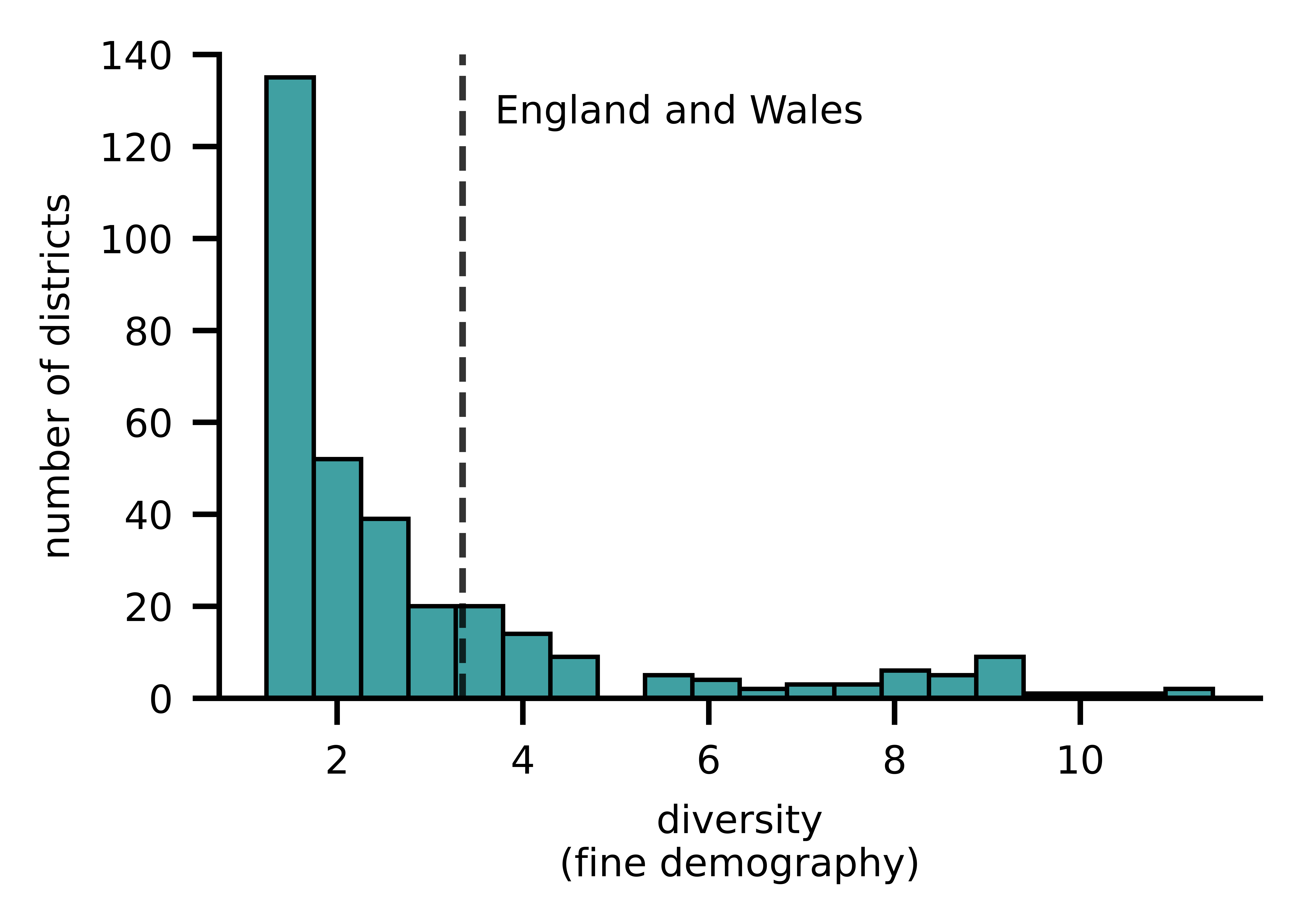}
    \includegraphics{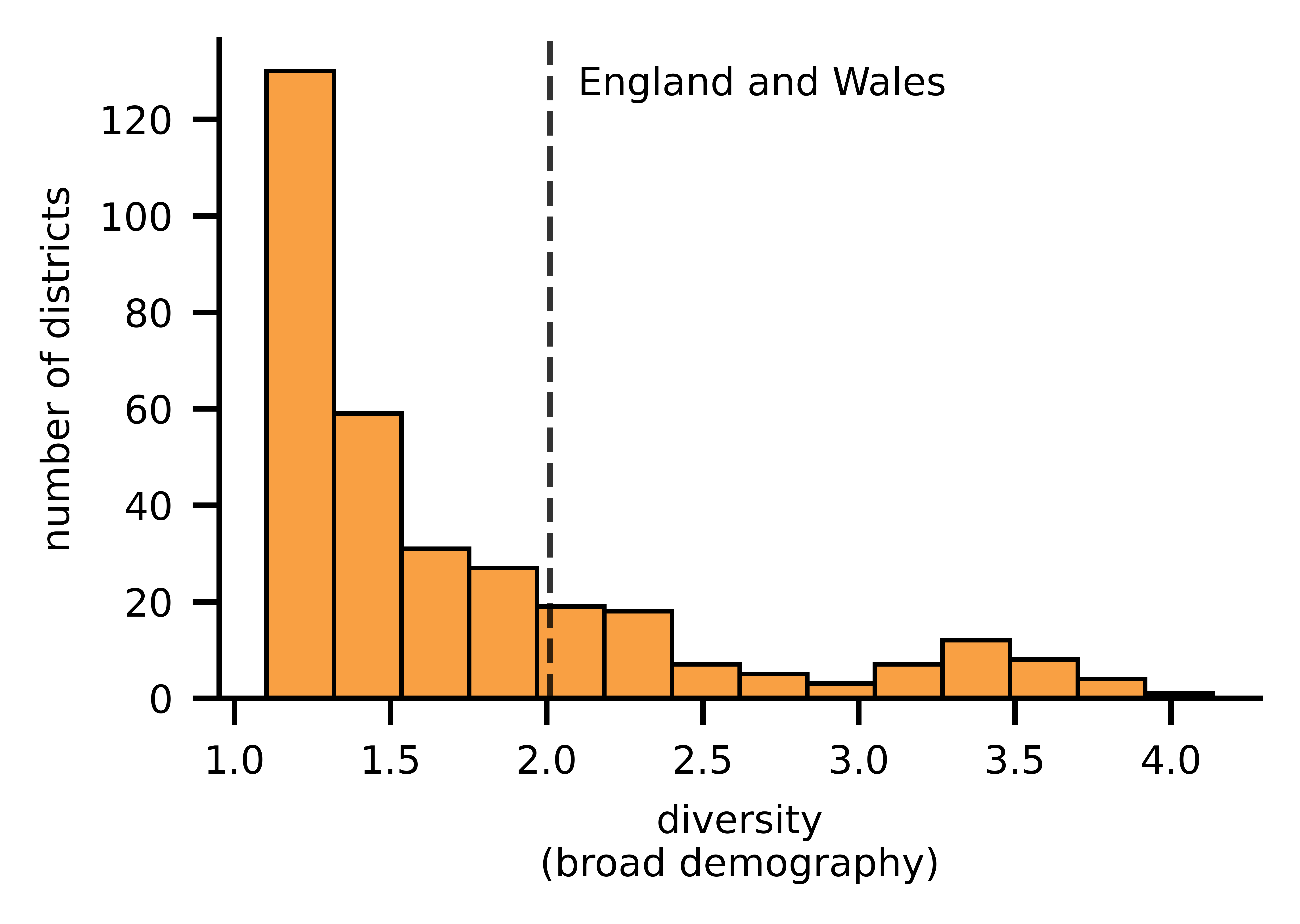}
    \caption{\textbf{Distribution of diversity in districts.} The diversities of districts at the fine (left, green) and broad (right, orange) demographic scales. The dashed line marks the overall diversity of England and Wales.}
\label{fig:dist_diversity_lads}
\end{figure}

\begin{figure}[h]
    \centering
    \includegraphics{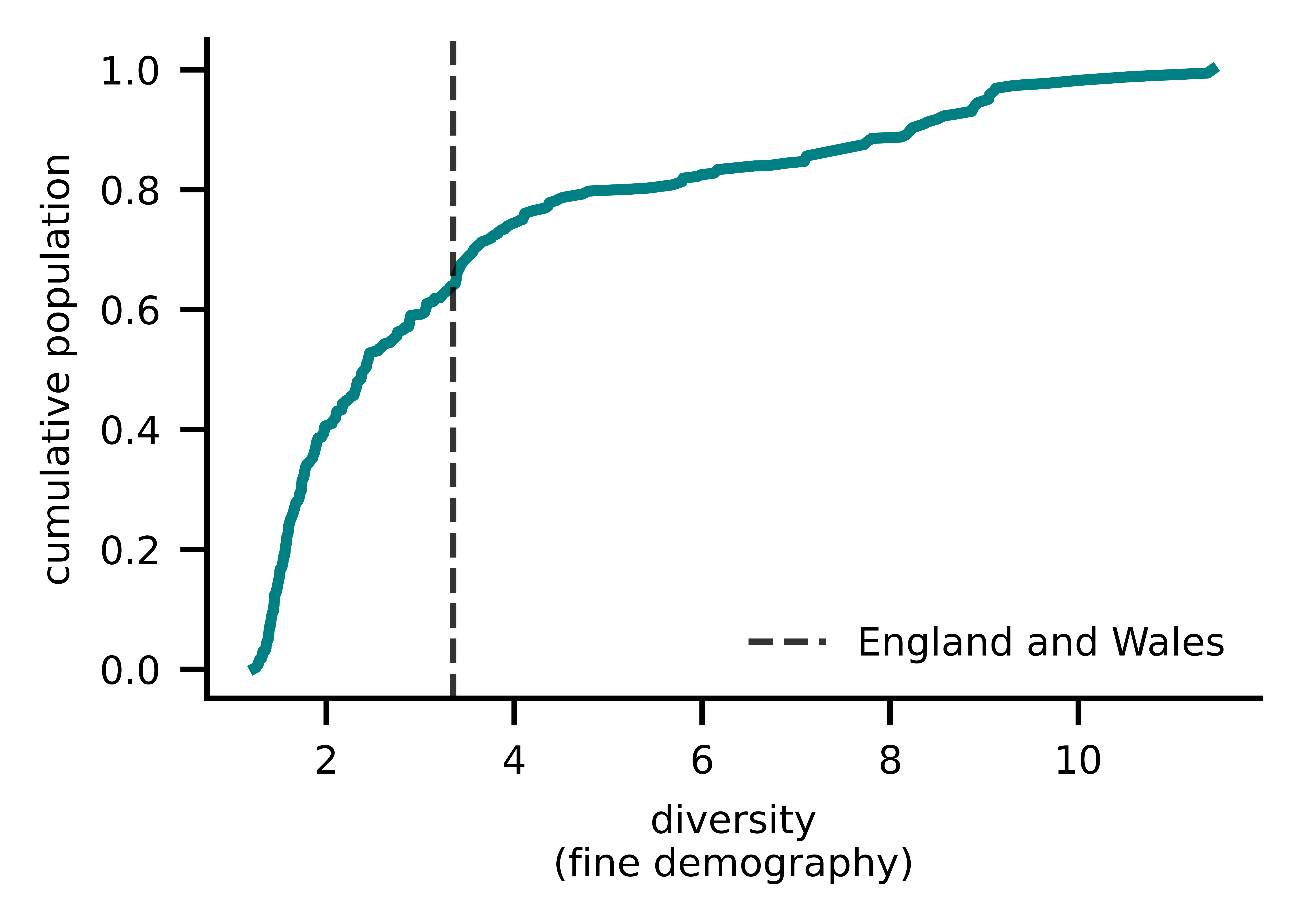}
    \includegraphics{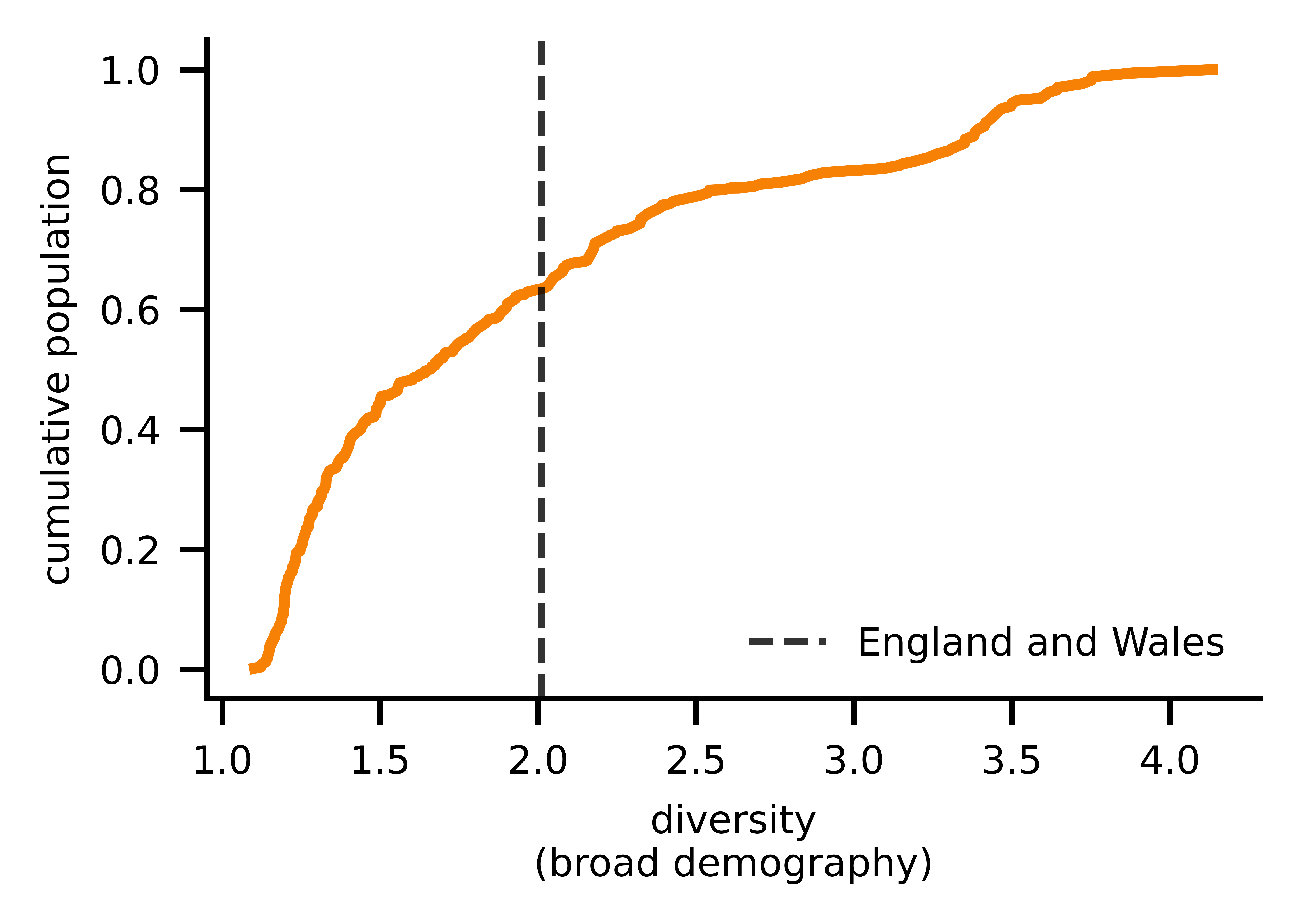}
    \caption{\textbf{Cumulative population vs diversity at district level.} Ordering the districts by diversity at the fine (left, green) and broad (right, orange) demographic scales, we plot the cumulative distribution of population. The dashed line marks the overall diversity of England and Wales.}
\label{fig:line_cumpop_vs_diversity_lads}
\end{figure}

\begin{figure}
    \centering
    \includegraphics{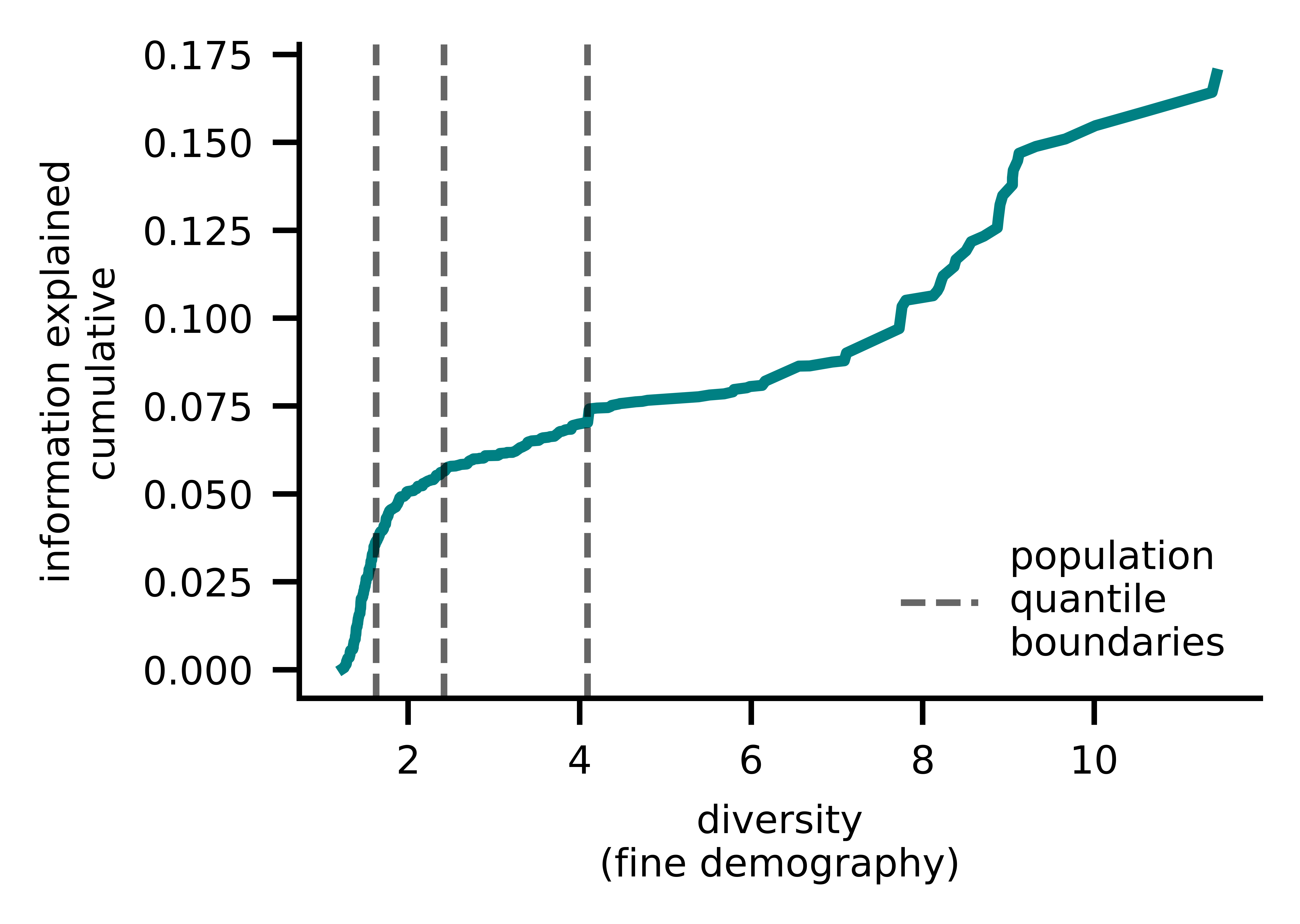}
    \includegraphics{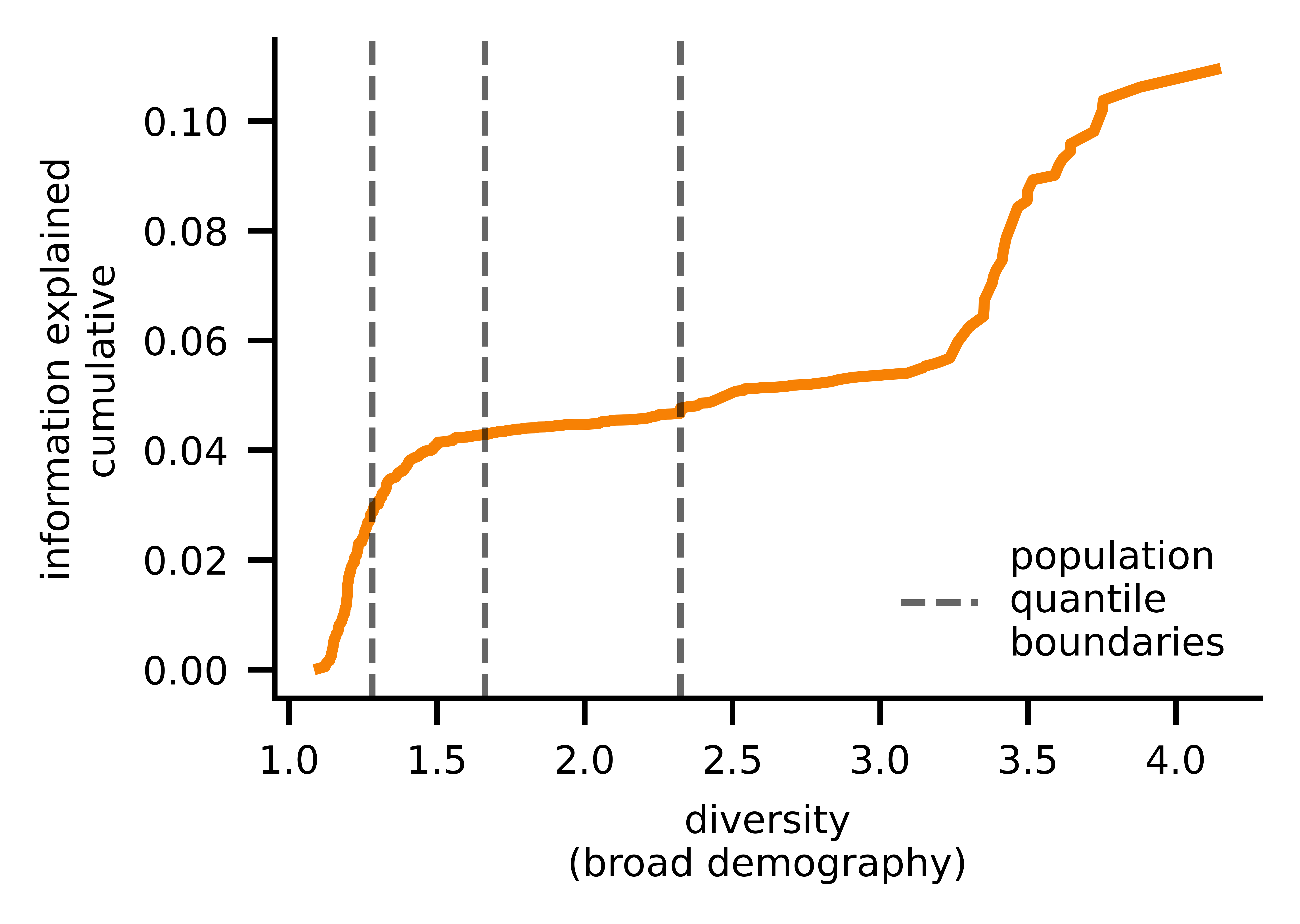}
    \caption{\textbf{Cumulative information explained against diversity at the district scale.} Ordering districts by increasing diversity, we plot the cumulative information explained (y-axis) against diversity (x-axis) at the fine (top, green) and coarse (bottom, orange) demographic scales. The dashed vertical lines denote the quantile boundaries of population. For instance, 25\% of the population lives in districts to the left of the first line.}
\label{fig:cuminfexplained_vs_diversity}
\end{figure}

\begin{figure}
    \centering
    \includegraphics{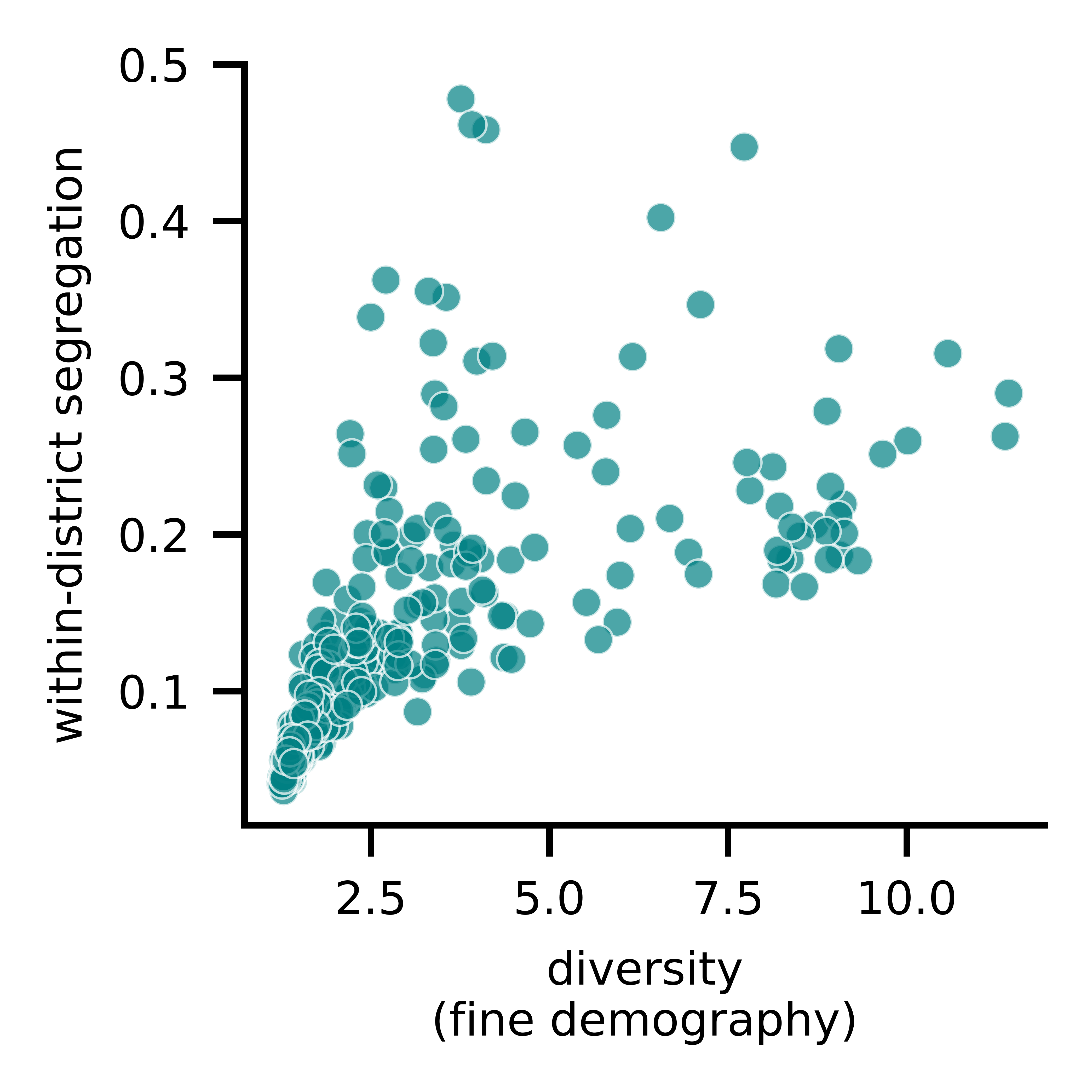}
    \caption{\textbf{Within-district segregation against diversity at the fine demographic scale.} Each point corresponds to a district.}
    \label{fig:scatter_segregation_vs_diversity_fine_lads}
\end{figure}

\begin{figure}
   \centering
        \includegraphics{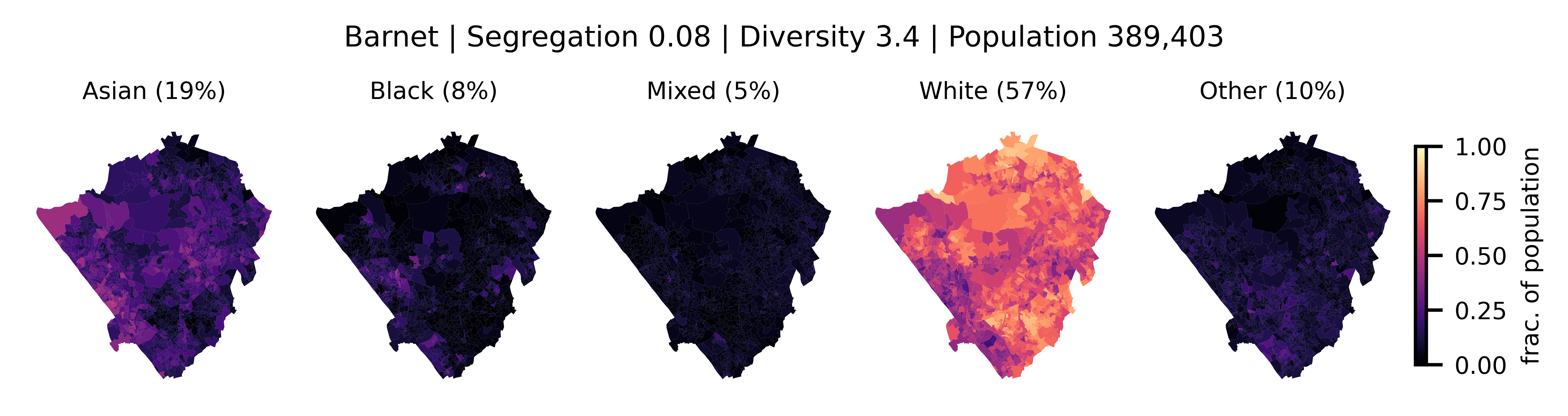}
        \includegraphics{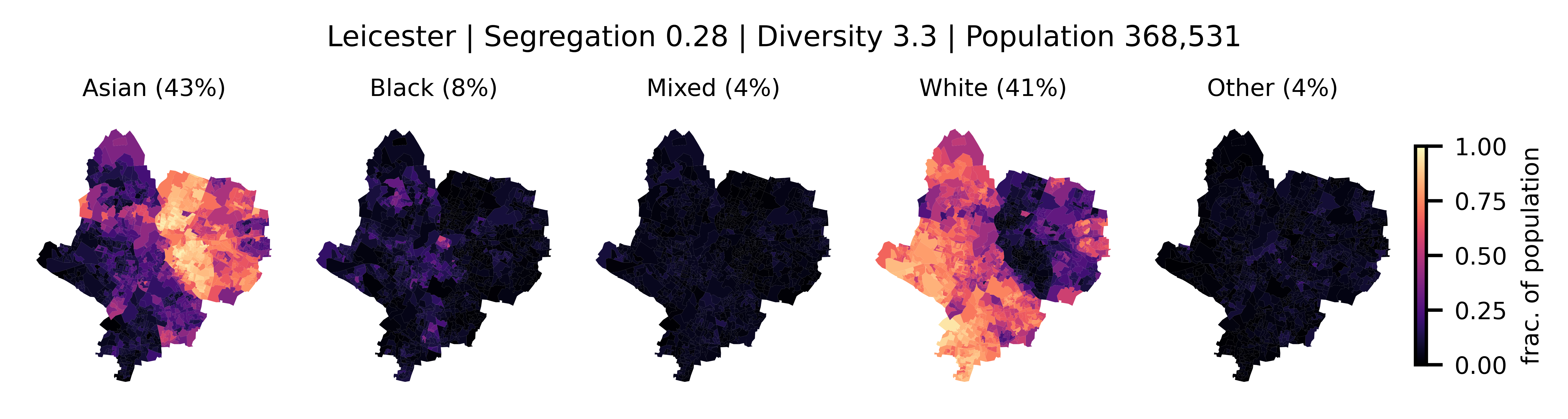}
    \caption{\textbf{Population distribution by ethnicity within Barnet and Leicester.} Each subplot depicts the distribution of a broad ethnic group within Barnet (top) and Leicester (bottom) at the OA scale. The colour of an OA indicates the fraction of its population that belongs to the ethnic group. While they have similar demographic compositions, the two districts show very different levels of segregation.}
\label{fig:maps_barnet_leicester}
\end{figure}

\begin{figure}
    \centering
    \includegraphics{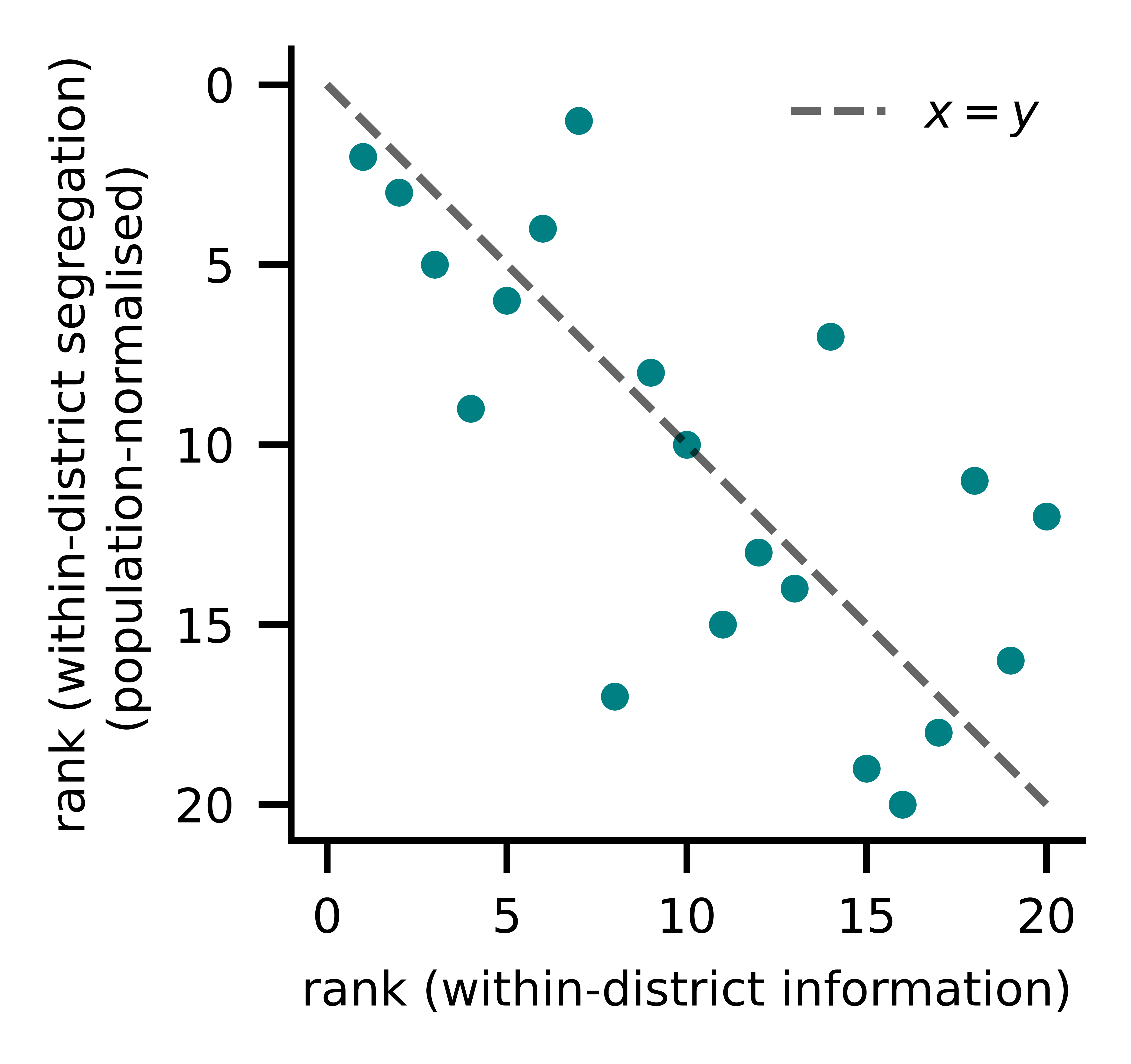}
    \includegraphics{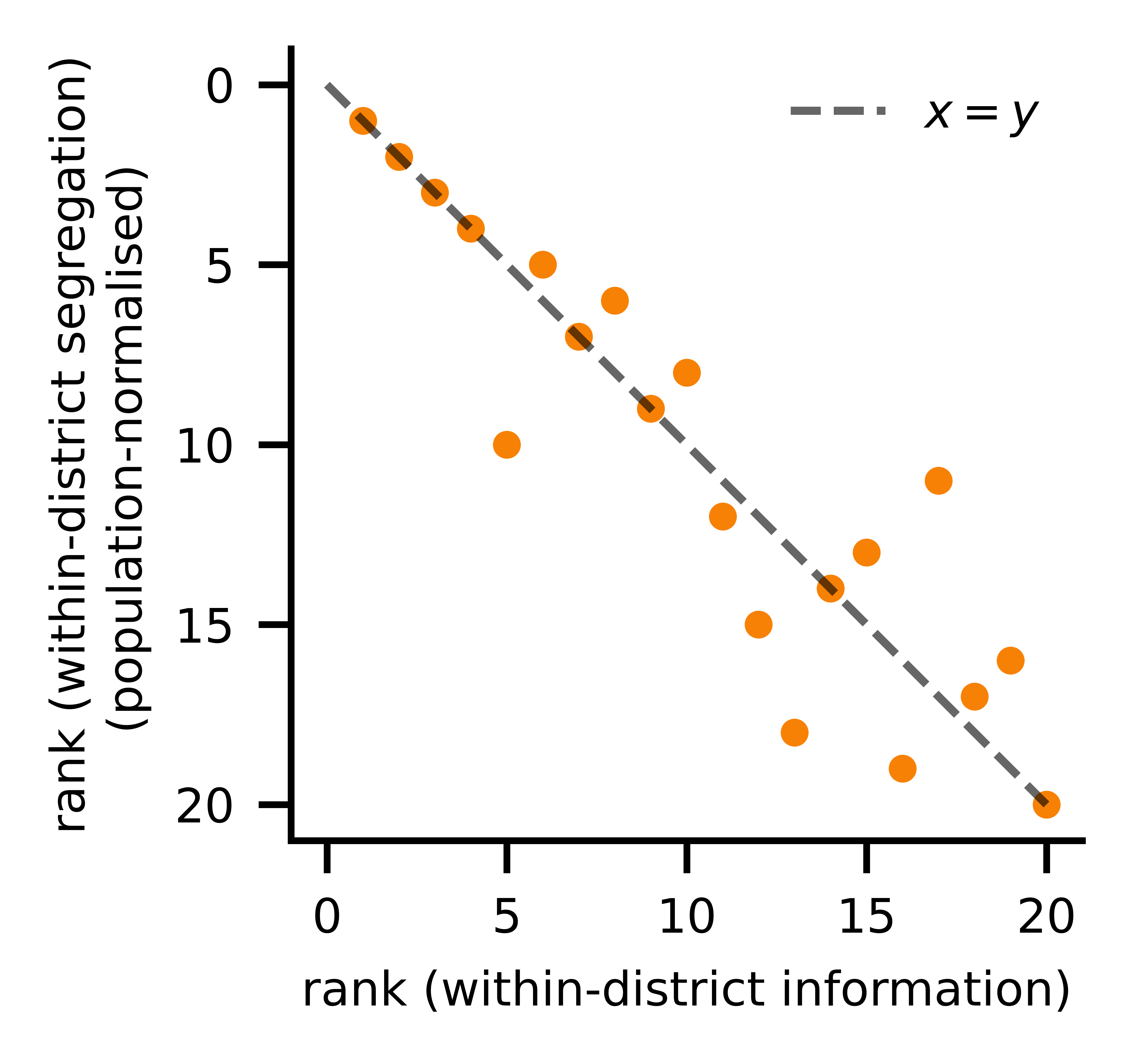}
    \caption{\textbf{Rankings by within-district information.} Rankings by $I_{\tilde{s}}^{\tilde{\mathcal{X}}}$ (y-axis) and $I_{\tilde{s}}^{\tilde{\mathcal{X}}}$ (x-axis) where a higher rank corresponds to more segregation. We plot the ranks of the 20 most diverse districts (ranked by the corresponding demographic scale) out of the 50\% most populous. The dashed line marks $y=x$. }
    \label{fig:scatter_rank_inf_vs_popnorm}
\end{figure}

\begin{figure}
    \centering
    \includegraphics{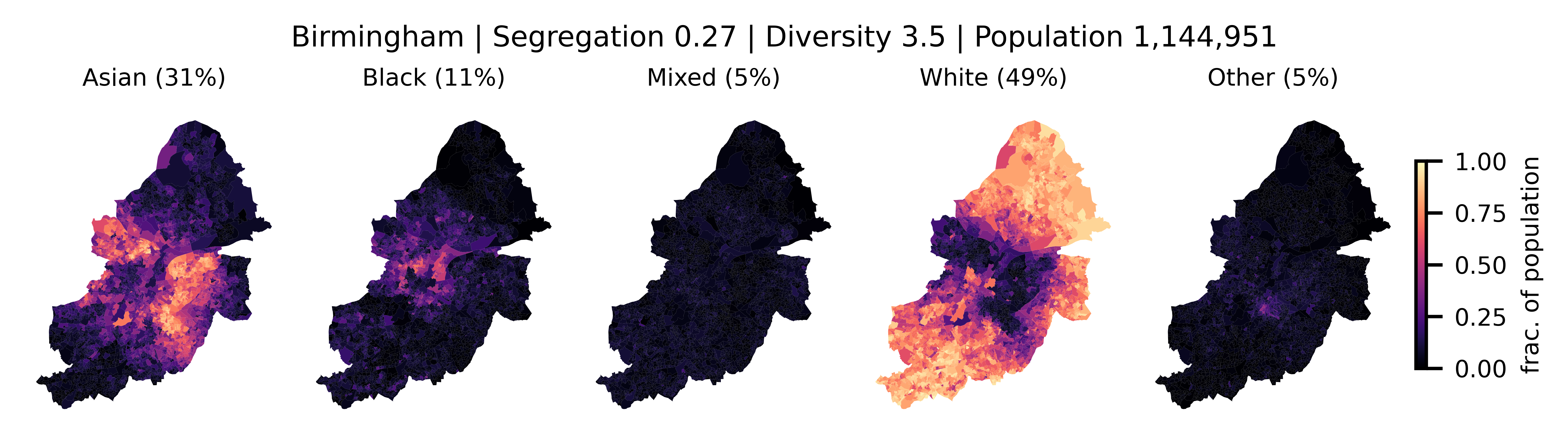}\\
    \includegraphics{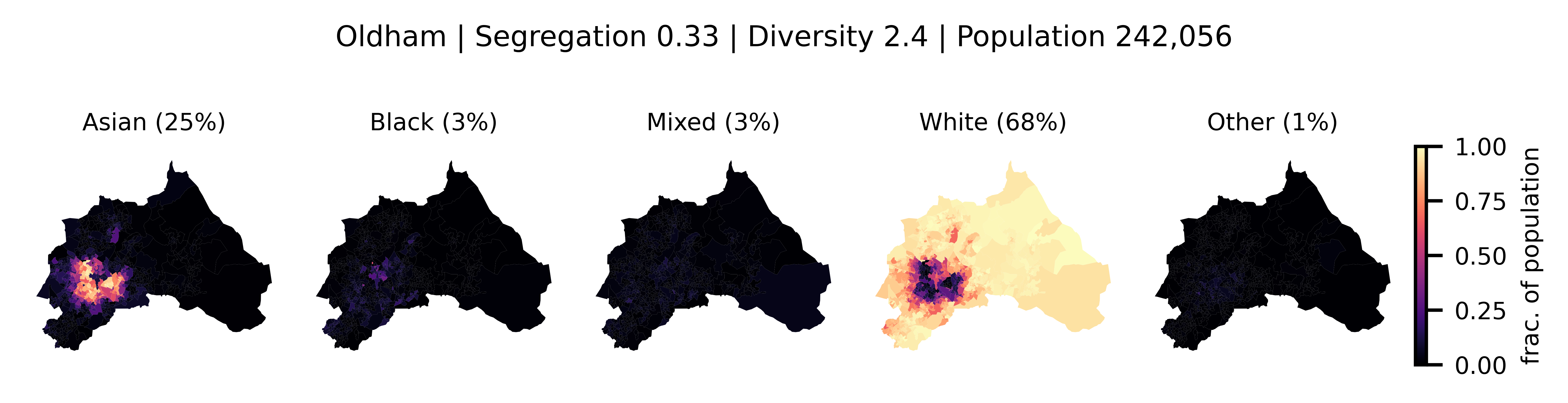}
    \caption{\textbf{Population distribution by ethnicity in Birmingham and Oldham.} Each subplot depicts the distribution of a broad ethnic group within Birmingham (top) and Oldham (bottom) at the OA scale. The colour of an OA indicates the fraction of its population that belongs to the ethnic group.}
    \label{fig:maps_birmingham_oldham}
\end{figure}

\begin{figure}
    \centering
    \includegraphics{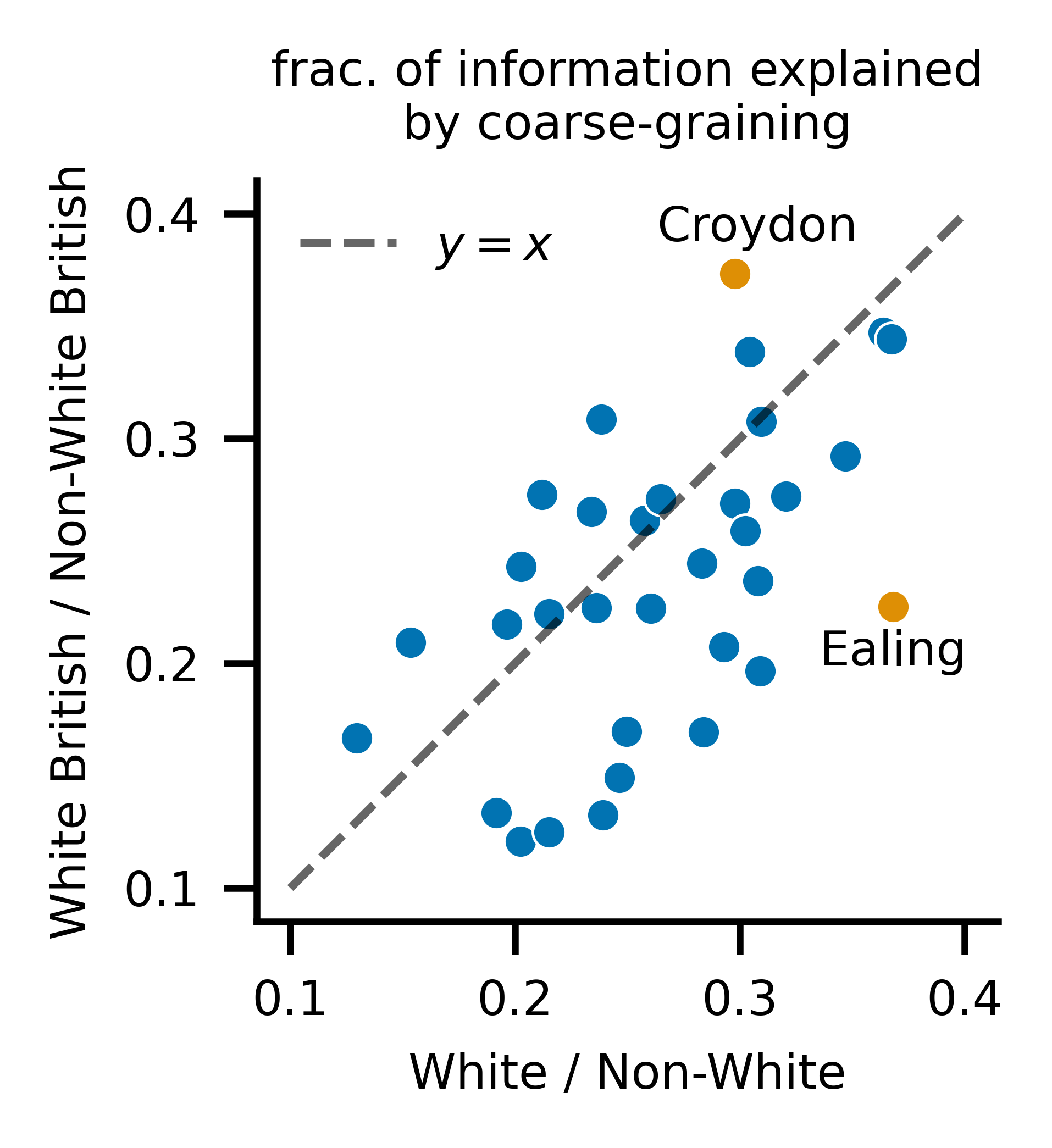}
    \caption{\textbf{Information explained by two-group demographic coarse-grainings.} The fraction of the total within-district information captured by the \textit{White} vs \textit{non-White} grouping (x-axis) and the \textit{White British} vs \textit{non-White British} grouping (y-axis) for the districts in Greater London. Croydon and Ealing are highlighted in orange and we visualise their population distribution in \cref{fig:maps_croydon_ealing}. Of the districts, the \textit{White British} vs \textit{non-White British} scale explains more information ($y>x$) for Barking and Dagenham, Barnet, Bexley, Bromley, Croydon, Enfield, Harrow, Havering, Kingston upon Thames, Richmond upon Thames, Sutton, and Waltham Forest. The opposite is true for City of London, Brent, Camden, Ealing, Greenwich, Hackney, Hammersmith and Fulham, Haringey, Hillingdon, Hounslow, Islington, Kensington and Chelsea, Lambeth, Lewisham, Merton, Newham, Redbridge, Southwark, Tower Hamlets, Wandsworth, and Westminster.}
    \label{fig:scatter_2group_coarsegraining_comparison}
\end{figure}

\begin{figure}
    \centering
    \includegraphics{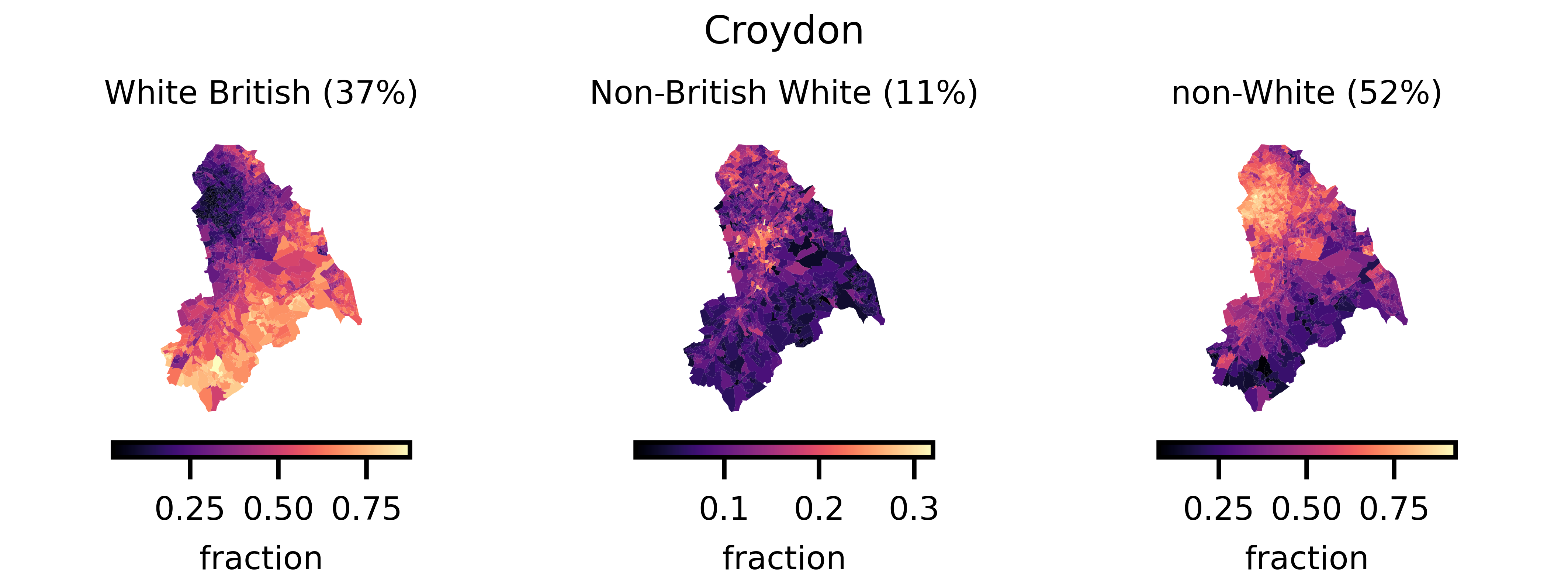}\\
    \includegraphics{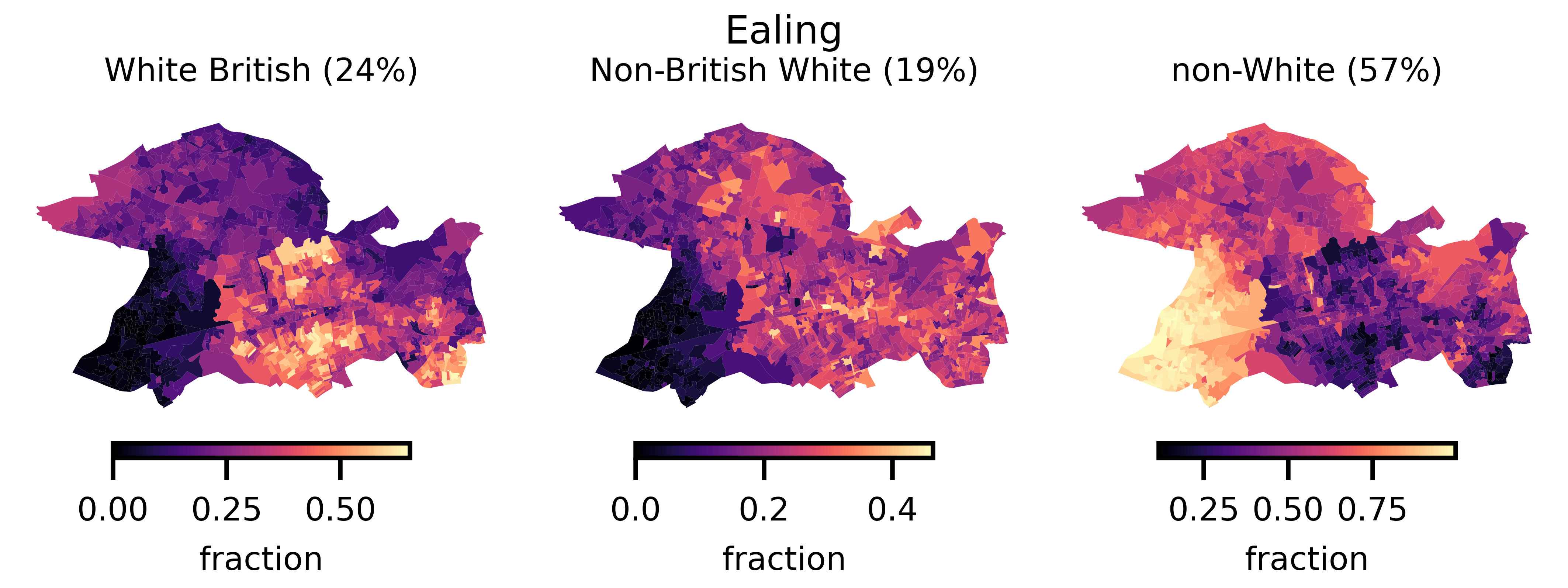}
    \caption{\textbf{Population distribution by ethnicity in Croydon and Ealing.} The distribution of \textit{White British}, \textit{non-British White}, and \textit{non-White} within Croydon (top) and Ealing (bottom) at the OA scale. The colour of an OA indicates the fraction of its population that belongs to the ethnic group.}
    \label{fig:maps_croydon_ealing}
\end{figure}

\end{document}